\documentclass[journal]{IEEEtran}
\usepackage{amsmath}
\usepackage{amsthm}
\newtheorem{lemma}{Lemma}
\usepackage{cancel}

\usepackage{amssymb}
\usepackage{amsfonts}
\usepackage{utfsym} 
\usepackage[colorlinks=true,urlcolor=black]{hyperref}
\usepackage{cleveref}
\crefname{section}{Sec.}{Secs.}
\Crefname{section}{Section}{Sections}
\Crefname{table}{Table}{Tables}
\crefname{table}{Tab.}{Tabs.}
\Crefname{figure}{Figure}{Figures}
\crefname{figure}{Fig.}{Figs.}
\Crefname{equation}{Equation}{Equations}
\crefname{equation}{Eq.}{Eqs.}
\crefname{algorithm}{Alg.}{Algs.}
\Crefname{algorithm}{Algorithm}{Algorithms}
\usepackage{url} 
\usepackage{cite} 
\usepackage{graphicx}
\usepackage{multirow}
\usepackage{makecell}
\usepackage[flushleft]{threeparttable}
\usepackage{colortbl}
\usepackage{booktabs}
\usepackage{caption}
\usepackage{subcaption}
\usepackage{xspace}
\usepackage{xcolor}
\usepackage{enumitem}
\usepackage[most]{tcolorbox} 
\usepackage{algorithm}
\usepackage{algpseudocode} 

\def\eg{\emph{e.g.,}\xspace}

\def\ie{\emph{i.e.,}\xspace}
\def\etal{\emph{et al.}\xspace}
\def\vs{\emph{vs.}\xspace}

\def\aka{\emph{a.k.a.}\xspace}

\newcommand{\one}{({\em i})\xspace}
\newcommand{\two}{({\em ii})\xspace}
\newcommand{\three}{({\em iii})\xspace}

\newcommand{\revise}[1]{\textcolor{black}{#1}}
\newenvironment{reviseG}
  {\begingroup\color{black}}
  {\endgroup}

\tcbset{
  takeaway/.style={
    colback=gray!10,       
    colframe=gray!60,      
    boxrule=0.8pt,         
    arc=4mm,               
    left=5mm, right=5mm,   
    top=3mm, bottom=3mm,   
    fonttitle=\bfseries,
    coltitle=black
  }
}

\ifCLASSINFOpdf
\else
\fi
\begin{document}
%
\title{Minimal Cascade Gradient Smoothing for\\Fast Transferable Preemptive Adversarial Defense}
%
%
%

\author{Hanrui Wang$^*$,~
        Ching-Chun Chang,~\IEEEmembership{Senior~Member,~IEEE,}
        Chun-Shien Lu,~
        Ching-Chia Kao,~
        \\
        Shuo Wang,~\IEEEmembership{Senior~Member,~IEEE,}
        and Isao Echizen,~\IEEEmembership{Senior~Member,~IEEE,}
\thanks{$^*$Corresponding author.}
\thanks{Hanrui Wang, Ching-Chun Chang, and Isao Echizen are with Echizen Lab, National Institute of Informatics (NII), Tokyo, Japan, e-mail: \{hanrui\_wang, ccchang, iechizen\}@nii.ac.jp. Chun-Shien Lu and Ching-Chia Kao are with Institute of Information Science, Academia Sinica, Taipei, Taiwan, e-mail: \{lcs, cck123\}@iis.sinica.edu.tw. Shuo Wang is with Shanghai Jiao Tong University, Shanghai, China, e-mail: wangshuosj@sjtu.edu.cn.}
\thanks{This work was partially supported by JSPS KAKENHI Grants JP21H04907 and JP24H00732, by JST CREST Grant JPMJCR20D3 and JPMJCR2562 including AIP challenge program, by JST AIP Acceleration Grant JPMJCR24U3, and by JST K Program Grant JPMJKP24C2 Japan.}
}

%
%

\markboth{IEEE Transactions on Information Forensics and Security,~Vol.~X, No.~X, Month~2026}%
{Wang \MakeLowercase{\textit{et al.}}: Minimal Sufficient Preemptive Defense}
%



\maketitle

\begin{abstract}
Adversarial attacks persist as a major challenge in deep learning. While training- and test-time defenses are well-studied, they often reduce clean accuracy, incur high cost, or fail under adaptive threats. In contrast, preemptive defenses, which perturb media before release, offer a practical alternative but remain slow, model-coupled, and brittle.
We propose the \textbf{Minimal Sufficient Preemptive Defense (MSPD)}, a fast, transferable framework that defends against future attacks without access to the target model or gradients. MSPD is driven by \textbf{Minimal Cascade Gradient Smoothing (MCGS)}, a two-epoch optimization paradigm executed on a surrogate backbone. This defines a minimal yet effective regime for robust generalization across unseen models and attacks.
MSPD runs at 0.02s/image (CIFAR-10) and 0.26s/image (ImageNet), 28--1696× faster than prior preemptive methods, while improving robust accuracy by +5\% and clean accuracy by +3.7\% across 11 models and 7 attacks.
To evaluate adaptive robustness, we introduce \textbf{Preemptive Reversion}, the first white-box diagnostic attack that cancels preemptive perturbations under full gradient access. Even in this setting, MSPD retains a +2.2\% robustness margin over the baseline. In practice, when gradients are unavailable, MSPD remains reliable and efficient. MSPD, MCGS, and Preemptive Reversion are each supported by \textbf{formal theoretical proofs}.
\textit{The implementation is available at \url{https://github.com/azrealwang/MSPD}.}
\end{abstract}


%
\IEEEpeerreviewmaketitle

\section{Introduction}
\label{intro}
Deep learning models are susceptible to adversarial attacks \cite{goodfellow2014explaining}, which introduce imperceptible perturbations to media content such as images, audio, or video. These perturbations can cause facial images to evade recognition~\cite{wang2021similarity,wang2024amulti} or mislead autonomous systems by altering traffic signs~\cite{eykholt2018robust}. Despite substantial progress, robust defense remains an open challenge.

\textbf{Training-time defenses}, \aka adversarial training~\cite{goodfellow2014explaining,shafahi2019adversarial,wong2020fast,chen2020adversarial,rade2021helper,addepalli2022efficient,sehwag2022robust,debenedetti2023light,xue2023xploring,wang2023better,peng2023robust,bartoldson2024adversarial,liu2024comprehensive,gowal2021improving,singh2023revisiting}, often face a trade-off between robustness and clean accuracy~\cite{zhang2019theoretically}. Fast variants reduce cost~\cite{wong2020fast} but typically yield weaker robustness than methods leveraging large auxiliary datasets or complex architectures~\cite{peng2023robust,bartoldson2024adversarial}.
To improve flexibility, \textbf{test-time defenses} such as adversarial detection~\cite{cohen2020detecting,zuo2021Exploiting,liang2021Detecting,hickling2023robust} and purification~\cite{samangouei2018defense,guo2018countering,raff2019barrage,shi2021online,yoon2021adversarial,hill2021stochastic,nie2022diffusion,yang2024adversarial,song2024mimicdiffusion,bai2024diffusion,chen2024diffilter,tang2024robust,wang2025iterative,zollicoffer2025lorid,lei2025instant,li2025adbm} offer plug-and-play alternatives. However, they incur inference overhead and reduce clean accuracy due to false positives or content distortion. Even state-of-the-art (SOTA) diffusion purifiers~\cite{nie2022diffusion} remain vulnerable to adaptive attacks~\cite{athalye2018obfuscated,xue2023diffusion,kang2023diffattack,wang2024diffhammer}.

\begin{figure}[!t]
    \centering
    \includegraphics[width=\linewidth]{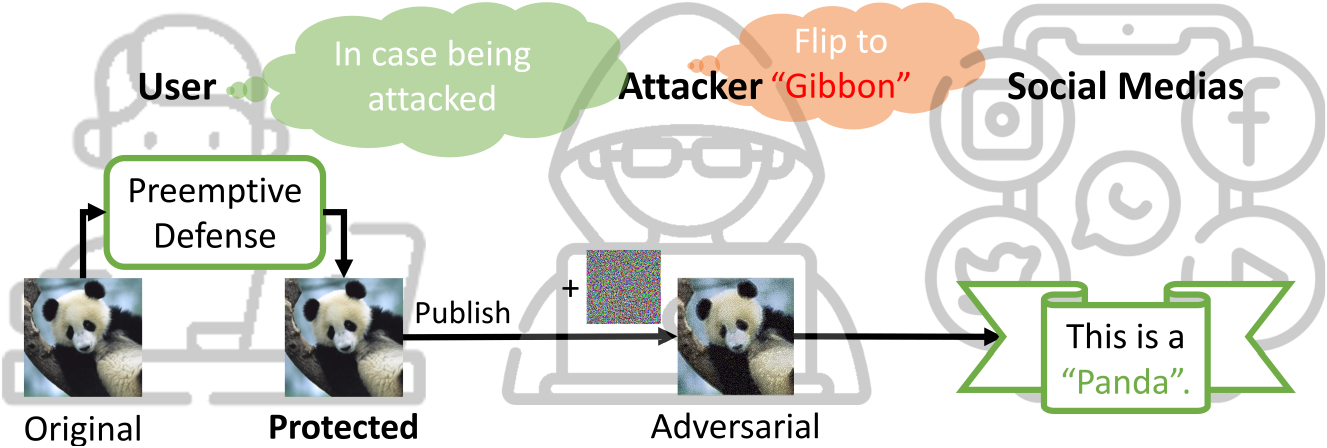}
    \caption{Overview of our preemptive defense. User images are proactively protected by embedding perturbations that anticipate and neutralize likely unseen attacks. When an actual attack occurs, these preemptive perturbations cancel the adversarial noise. The defense is transferable and generalizes well across unseen models without customization.}
    \label{fig_application}
\end{figure}

Recent works have explored \textbf{preemptive defenses} \cite{salman2021unadversarial,moon2022preemptive,frosio2023best}, which proactively embed protective perturbations \emph{before} media release to defend against unknown future attacks. As illustrated in \cref{fig_application}, this approach enables one-time protection that generalizes across unseen threats and platforms (\eg social media) without target-specific tuning. This matches common creator workflows, local processing on phones or desktops before upload. Industry practices like watermarking and C2PA tags~\cite{rosenthol2022c2pa} and on-device filters reflect this norm.

Despite their promise, existing preemptive defenses face key \textbf{limitations:} \textbf{\one High processing cost:} inefficient algorithms impose significant overhead per image. \textbf{\two Limited generalizability:} many methods overfit to specific backbones, reducing robustness on unseen or mismatched models (\eg trained on Facebook but used on Instagram). \textbf{\three Lack of adaptive threat modeling:} most assume static attackers and remain vulnerable when the defense is known. The emerging paradigm is gaining traction, with applications in face recognition, deepfake prevention, and uncopyrighted generation~\cite{li2025defense,mi2025preemptive,gui2025i2vguard}, motivating research into scalability and robustness.

We propose the \textbf{Minimal Sufficient Preemptive Defense (MSPD)}, a low-latency, transferable defense framework that proactively embeds lightweight perturbations before media release, addressing three persistent challenges in adversarial robustness: efficiency, generalization to unseen models and attacks, and resilience to adaptive threats. Unlike adversarial training, MSPD requires no access to target models, gradients, or attack parameters. It identifies shared vulnerabilities, input regions consistently exploited across diverse attacks and models~\cite{tramer2018ensemble,wang2024diffhammer}, and injects directionally reversed perturbations that neutralize these weak points without target-specific tuning. Its fully decoupled design allows users to configure the defense privately, like a secret key, enhancing robustness against strong adaptive adversaries.

At the core of MSPD is the \textbf{Minimal Cascade Gradient Smoothing (MCGS)} algorithm, a two-epoch learning procedure that proves to be both computationally minimal and empirically sufficient. Our key insight is that a two-epoch cascade, applying single-step gradient smoothing in both forward and backward passes, is sufficient to achieve near-optimal robustness. The forward step pushes inputs away from decision boundaries to reduce gradient sensitivity; the backward step reverses likely attack directions to enhance generalization; and smoothing in each step ensures that perturbations align with robust, transferable subspaces. Despite its simplicity, MCGS achieves +5\% robust accuracy gain over the prior SOTA~\cite{moon2022preemptive}, across 11 unseen models and 7 attacks, while being 28--1696$\times$ faster (0.02–0.26s/image). We further prove that MCGS is not a trivial blend of forward- and backward-only baselines, but a distinct, theoretically grounded strategy that delivers strong clean and robust performance even under adaptive threats.

To assess resilience under image processing and adaptive threats, we test if our defense survives JPEG compression and diffusion-based purification, which may disrupt preemptive perturbations. We find these transformations cause only distortion without neutralizing the defense. We therefore introduce \textbf{Preemptive Reversion}, a stronger adaptive diagnostic (stress test) not limited to our MSPD, which shows that a secondary defense with the same backbone approximates similar perturbations despite randomness. Results show that only an attacker with full access to both the defense budget and backbone gradients (an unrealistic white-box assumption) can partially reverse the preemptive defense. Even in this setting, MSPD retains \textbf{+2.2\%} higher robustness than the undefended baseline, demonstrating resilience to strong adaptive adversaries. \textit{Note:} The only feasible adaptive strategy is to purify the protected image (we reverse it). Post-hoc attacks like BPDA+EOT do not neutralize the preemptive signal, while adaptive training or detection are impractical since the defense operates locally, outside the attacker's control. Even if the defense is known, the attacker cannot disrupt this user-side protection, but can only attempt to remove the added perturbation afterward.

Our key insight is that robustness can be achieved through the simple known configuration effective against adaptive, black-box, and white-box threats, empirically near-minimal computational cost for preemptive defenses.
We summarize our key contributions as follows:
\begin{enumerate}
    \item \textbf{MSPD:} A low-latency, transferable preemptive defense that improves clean and robust accuracy while protecting against unseen models and attacks.
    \item \textbf{MCGS:} A two-epoch learning algorithm that applies single-step gradient smoothing in both forward and backward passes, achieving +5\% robustness and 28--1696$\times$ speedup over prior SOTA.
    \item \textbf{Preemptive Reversion:} A white-box diagnostic that counteracts preemptive perturbations under full gradient access, enabling stress test of adaptive robustness.
    \item \textbf{Theoretical Foundations:} Four formally proven lemmas
    theoretically justify MSPD's design, supporting its robustness, generalization, and convergence under minimal configuration.
\end{enumerate}

\begin{table}[!t]
    \scriptsize
    \centering
    \begin{threeparttable}
        \caption{Overview of related works on preemptive defenses.}
        \label{tab_related_works}
        \setlength{\tabcolsep}{1.6mm}{\begin{tabular}{lccccc}
            \toprule
            Method&Efficient$^1$&\begin{tabular}{c}Unseen\\Model\end{tabular}&\begin{tabular}{c}Unseen\\Attack\end{tabular}&\begin{tabular}{c}Adaptive\\Bench.\end{tabular}&\begin{tabular}{c}Code\\Avail.\end{tabular}\\
            \midrule
            UnAdv~\cite{salman2021unadversarial}&$\usym{2717}$&$\usym{2717}$&$\usym{2713}$&$\usym{2717}$&$\usym{2713}$\\
            Bi-Level~\cite{moon2022preemptive}&$\usym{2717}$&$\usym{2713}$&$\usym{2713}$&$\usym{2717}$&$\usym{2713}$\\
            A$^5$~\cite{frosio2023best}&$\usym{2713}$&$\usym{2717}$&$\usym{2713}$&$\usym{2717}$&$\usym{2713}$\\
            SGSD~\cite{li2025defense}&$\usym{2713}$&$\usym{2717}$&$\usym{2713}$&$\usym{2717}$&$\usym{2717}$\\
            BBPD~\cite{mi2025preemptive}&$\usym{2717}$&$\usym{2717}$&$\usym{2713}$&$\usym{2717}$&$\usym{2717}$\\
            I2VGuard~\cite{gui2025i2vguard}&$\usym{2717}$&$\usym{2717}$&$\usym{2717}$&$\usym{2717}$&$\usym{2717}$\\
            \rowcolor{gray!10}
            \textbf{MSPD (Ours)}&$\usym{2713}$&$\usym{2713}$&$\usym{2713}$&$\usym{2713}$&$\usym{2713}$\\
            \bottomrule
        \end{tabular}}
        \begin{tablenotes}
            \item $^1$ A$^5$, SGSD, and our defense complete protection within $2$ seconds on images size of $224 \times 224$ to $256 \times 256$, whereas UnAdv, Bi-Level, BBPD, and I2VGuard require 38--150 seconds.
        \end{tablenotes}
    \end{threeparttable}
\end{table}

\begin{table*}[!t]
\scriptsize
\centering
\caption{Comparison of preemptive, training-time, and test-time defenses.}
\label{tab_defense_families}
\setlength{\tabcolsep}{1.1mm}{\begin{tabular}{l|l|l|l|l}
\toprule
Defense&Expectation&Pros&Cons&Application\\
\midrule
  \begin{tabular}{l}
    Preemptive\\\textbf{(Ours)}
  \end{tabular}
  &\begin{tabular}{l}
    Orig. images are protected
  \end{tabular}
  &\begin{tabular}{l}
    + Highest robustness\\
    + Test-time efficient
  \end{tabular}
  &\begin{tabular}{l}
    - Inapplicable for nonrobust models\\ 
    - Only protects perturbed exemplars
  \end{tabular} 
  &\begin{tabular}{l}
    End-user side deployment\\ 
    Face recognition apps\\ 
    Deepfake prevention
  \end{tabular}\\
\midrule
  \begin{tabular}{l}
    Training-Time
  \end{tabular}
  &\begin{tabular}{l}
    Adv. examples can be\\ 
    classified correctly
  \end{tabular}
  &\begin{tabular}{l}
    + Test-time efficient\\ 
    + Improves clean accuracy over\\test-time defenses
  \end{tabular} 
  &\begin{tabular}{l}
    - Mostly sacrifices clean accuracy\\ 
    - Requires replacement of current models
  \end{tabular} 
  &\begin{tabular}{l}
    Standard supervised learning pipelines\\ 
    Cloud-based recognition models
  \end{tabular}\\
\midrule
  \begin{tabular}{l}
    Test-Time
  \end{tabular}
  &\begin{tabular}{l}
    Adv. examples can be\\ 
    purified or detected
  \end{tabular}
  &\begin{tabular}{l}
    + Applicable for nonrobust models\\ 
    + Plug-and-play implementation
  \end{tabular} 
  &\begin{tabular}{l}
    - Sacrifices clean accuracy\\ 
    - Adds test-time cost
  \end{tabular} 
  &\begin{tabular}{l}
    Third-party inference APIs\\ 
    Legacy models without retraining
  \end{tabular}\\
\bottomrule
\end{tabular}}
\end{table*}

\section{Related Works}

\subsection{Preemptive Defenses}

We review prior preemptive defenses, focusing on computational efficiency and generalization to unseen models and attacks (\cref{tab_related_works}). This defense family remains underexplored due to limited transferability and high cost. The closest baseline is Bi-Level~\cite{moon2022preemptive}, which degrades on unseen models and requires heavy training. UnAdversarial (UnAdv)~\cite{salman2021unadversarial} uses forward-only perturbations for physical corruption robustness but targets white-box settings and lacks adversarial evaluation; we adapt it as a forward-only digital baseline. Frosio~\etal~\cite{frosio2023best} propose a generator-based forward defense for physical attacks, but it struggles under tight perturbation budgets.

Recent work extends preemptive defenses to face recognition~\cite{li2025defense,mi2025preemptive} and generative pipelines~\cite{gui2025i2vguard}. However, methods such as I2VGuard~\cite{gui2025i2vguard} remain computationally heavy and exhibit limited transferability or adaptive robustness. These developments underscore the need for more generalizable and efficient preemptive defenses. The emergence of in-camera anti-deepfake filters further demonstrates the practical relevance of user-side perturbations.

\begin{figure}[!t]
    \centering
    \includegraphics[width=3.3in]{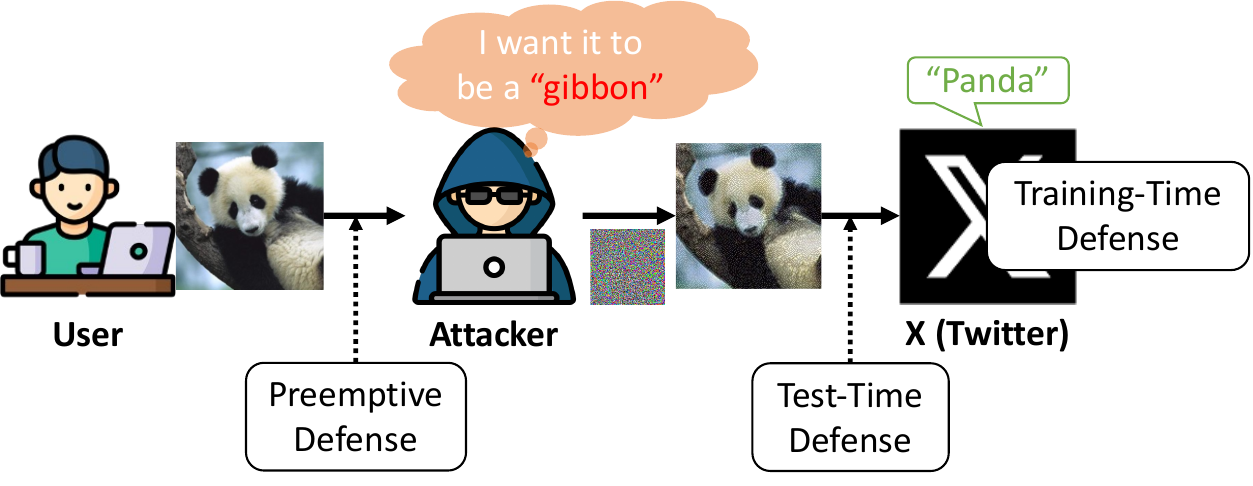}
    \caption{Defense families in the attack–defense pipeline: preemptive (ours), training-time, and test-time.}
    \label{fig_defense_families}
\end{figure}

\subsection{Comparison with Other Defenses}

Adversarial defenses fall into three categories: preemptive, training-time, and test-time (\cref{fig_defense_families}, \cref{tab_defense_families}). \textbf{Preemptive defenses} act \emph{before} model querying by perturbing inputs at creation time. They are test-time efficient and can improve clean accuracy, but cannot protect nonrobust backbones and only secure preemptively processed inputs. Applications include secure face recognition and generative provenance. \textbf{Training-time defenses} incorporate robustness into model parameters (e.g., adversarial training~\cite{madry2018towards}, TRADES~\cite{zhang2019theoretically}). They are inference-efficient and offer strong robustness–accuracy trade-offs, but require retraining and may reduce clean accuracy. \textbf{Test-time defenses} purify or detect adversarial inputs before inference (e.g., Defense-GAN~\cite{samangouei2018defense}, DiffPure~\cite{nie2022diffusion}). They are plug-and-play for pretrained models but incur additional inference cost and may degrade clean accuracy.

In summary, preemptive defenses protect at content creation, training-time defenses harden models through retraining, and test-time defenses retrofit robustness at inference. \revise{These complementary strategies can be layered to enhance robustness across the full attack-defense pipeline.}

\section{Minimal Sufficient Preemptive Defense}
\label{MSPD}
\begin{figure}[!t]
    \centering
    \includegraphics[width=2.6in]{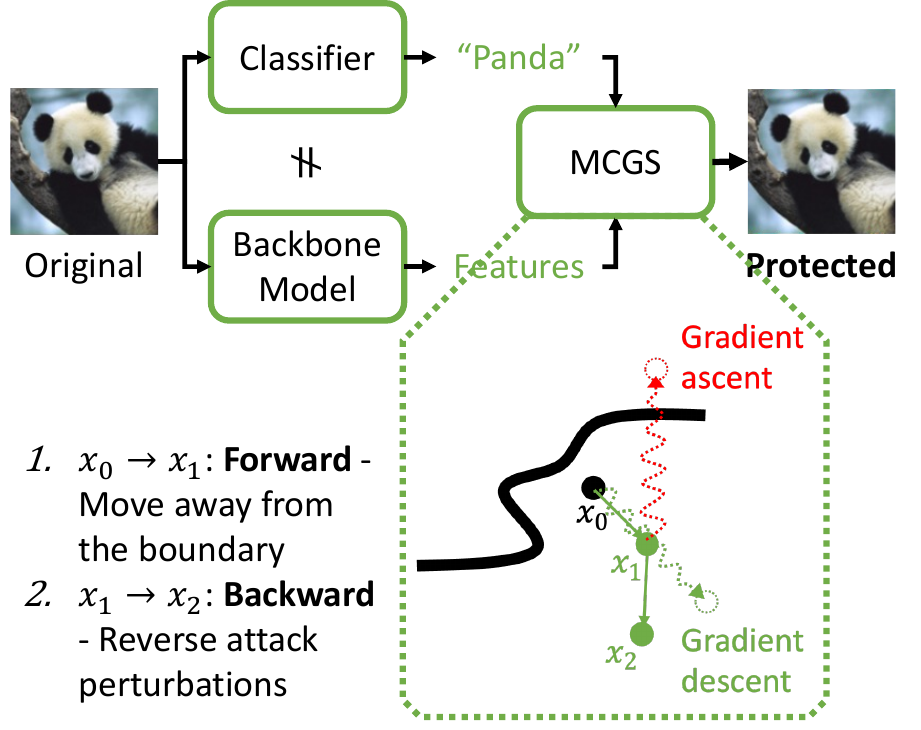}
    \caption{Overview of our preemptive defense pipeline. Unlike prior methods that couple components (\eg backbone = target or classifier), we decouple all roles (\textbf{classifier $\neq$ backbone $\neq$ target}) to improve generalizability and support realistic, user-side deployment. This design enables private ``secret-key" backbones, making gradient leakage unlikely. We further introduce MCGS, which converges in just \textbf{one forward and one backward epoch} using single-step smoothing, greatly reducing computation.}
    \label{fig_ourdefense}
\end{figure}

We first define the threat model in \cref{threat_model} and provide an overview of the defense pipeline in \cref{fig_ourdefense}. We then describe the backbone model (\cref{backbone_model}), classifier (\cref{classifier}), and learning algorithm (\cref{mcgs}).

\subsection{Threat Model}
\label{threat_model}
We define our threat model by outlining the roles and assumptions illustrated in \cref{fig_application}.

\textbf{User (Defender):}
The user applies the preemptive defense proactively before releasing media, mirroring standard creator workflows where the final editable copy resides on the user's device. The defense is trained without access to the attack algorithm, its parameters, or the architecture or gradients of downstream models. The goal is to defend against a broad range of \textbf{unseen attacks} on \textbf{unseen models}, emphasizing generalizability and transferable deployment.

\revise{\textbf{Attacker:}
We consider three knowledge and capability:
\begin{itemize}
    \item \textbf{White-Box:} Full access to the target model’s architecture and gradients, but unaware of the defense.
    \item \textbf{Black-Box:} No gradient access, but can query the target model and observe outputs, or even no access to the target model or its identity, relying solely on transferability.
    \item \textbf{Adaptive:} Aware of the defense and actively attempts to reverse or bypass it.
\end{itemize}
We exclude ``supply-chain" attacks on cameras or software \textit{before} export, as full-privilege cases nullify creator-side defenses and need distinct system-level safeguards. Additionally, throughout this paper we consider only \emph{untargeted} attacks (aiming at misclassification), since untargeted attacks are generally stronger baselines for success rate / robustness evaluation, thus harder to defend against than targeted ones.}

\textbf{Social Media Platforms (Target Models):}
These represent downstream models processing the protected media. Our strength is that target models \textbf{require no changes or defense awareness}, avoiding retraining and deployment overhead, crucial for large-scale, resource-intensive systems.

\subsection{Backbone Model Selection}
\label{backbone_model}
In our threat model, the defender has no knowledge of the target models but aims to protect them in advance. To defend unseen targets using MSPD, the defender must select a single backbone $f_{\mathrm{back}}$ that is independent of the actual target model $f_{\mathrm{tgt}}$, thereby avoiding target-specific assumptions. The final goal is to enhance robustness across many unseen target models (\ie \emph{target-model robustness only}) by training on a single backbone, rather than maximizing robustness of the backbone itself (\eg via optimal convergence). This requires protection that is transferable among unseen targets. We show that effective transfer hinges on the alignment between gradients $\nabla_x f_{\mathrm{back}}(x)$ and $\nabla_x f_{\mathrm{tgt}}(x)$ within a shared vulnerable subspace. Stronger alignment yields a stronger protective effect, with the trivial case $f_{\mathrm{back}} = f_{\mathrm{tgt}}$ giving maximal alignment (\cref{tab_ablation_pipeline}) but falling outside our threat model. One of the key factors for successful defense is how to identify a backbone that is well aligned with many unseen models.

\begin{lemma}
\label{lemma_backbone}
Let $f_{\mathrm{back}}$ and $f_{\mathrm{tgt}}$ denote backbone and target models. If $\nabla_x f_{\mathrm{back}}(x)$ and $\nabla_x f_{\mathrm{tgt}}(x)$ share a vulnerable subspace, then defense perturbations $\delta^{\mathrm{def}}$ computed from $f_{\mathrm{back}}$ transfer to $f_{\mathrm{tgt}}$ and suppress future attacks.
\end{lemma}

\noindent
This condition favors robust backbones, which are known to yield semantically meaningful and architecture-agnostic gradients~\cite{tramer2018ensemble}. We adopt adversarially trained models from model zoos~\cite{croce2021robustbench}, as they empirically satisfy the lemma’s hypothesis and improve transfer across diverse architectures. The empirical evidence with success/failure case analysis are provided in \cref{tab_ablation_backbone} and \cref{tab_ablation_pipeline}. \revise{The proof is in Appendix~\ref{proof}.}

\subsection{Classifier Selection}
\label{classifier}
To guide preemptive perturbations, our framework employs a separate classifier $f_\mathrm{cls}$ to estimate labels. This classifier is \textit{decoupled} from the backbone $f_\mathrm{back}$ and is \textit{not} used for gradient computation. Its sole role is to provide the label $\hat{y} = \arg\max f_\mathrm{cls}(x)$ for loss supervision.
This design prioritizes label correctness by favoring clean accuracy over robustness. Unlike prior approaches that conflate the classifier and backbone, we find that decoupling yields stronger and more transferable defenses. The classifier architecture need not match the backbone or target model. As long as the predicted label satisfies \(\hat{y} = y\), the gradient direction remains semantically valid and aligned with the true task. This intuitive design choice is empirically validated through \cref{tab_ablation_pipeline}.

\subsection{Minimal Cascade Gradient Smoothing (MCGS)}
\label{mcgs}
We propose the MCGS algorithm, an efficient paradigm for estimating protective perturbations that generalize to unseen models and attacks. It consists of two key components: 1) a cascade of single-step forward (F) and backward (B) learning epochs, and 2) gradient smoothing within each epoch. MCGS converges in 2 epochs (F$\rightarrow$B) while matching or exceeding 100-epoch baselines; see \cref{tab_defense}.

\subsubsection{\textbf{Cascade Learning Strategy}}
\label{cascade}

Let $g_k = \nabla_x \mathcal{L}(x_k)$ denote the input gradient at step $k$ of a $K$-step adversarial training procedure using loss $\mathcal{L}$. At each step, the update is constrained by an $\ell_p$ projection $\Pi_{\epsilon}(\cdot)$ that ensures perturbations remain within the allowed budget $\epsilon$. We define two complementary update modes in MCGS (\cref{fig_ourdefense}):  
\begin{itemize}
    \item \textbf{Forward learning} applies projected gradient descent on the backbone’s cross-entropy loss with step size $\alpha > 0$:  
    \begin{equation}
        x_{k+1}^{\mathrm{F}} = \Pi_{\epsilon}\left(x_k^{\mathrm{F}} - \alpha g_k^{\mathrm{F}}\right), 
        \quad x_0^{\mathrm{F}} = x.
        \label{eq_forward}
    \end{equation}
    This moves the original input $x$ toward high-confidence regions, yielding fast convergence but prone to overfitting the backbone’s decision geometry.  
    
    \item \textbf{Backward learning} performs projected gradient ascent on the loss, followed by a reversal step:  
    \begin{equation}
      \begin{aligned}
        x_{k+1}^{\mathrm{adv}} = \Pi_{\epsilon}\left(x_k^{\mathrm{adv}} + \alpha g_k^{\mathrm{adv}}\right),\\ 
        x_0^{\mathrm{adv}} = x, \quad x_K^{\mathrm{B}} = 2x - x_K^{\mathrm{adv}}.
      \end{aligned}
      \label{eq_backward}
    \end{equation}
    This pushes the input away from error-prone regions without directly enforcing correctness, thereby enhancing generalization but converging more slowly.
\end{itemize}

\begin{figure}[!t]
    \centering
    \includegraphics[width=2.3in]{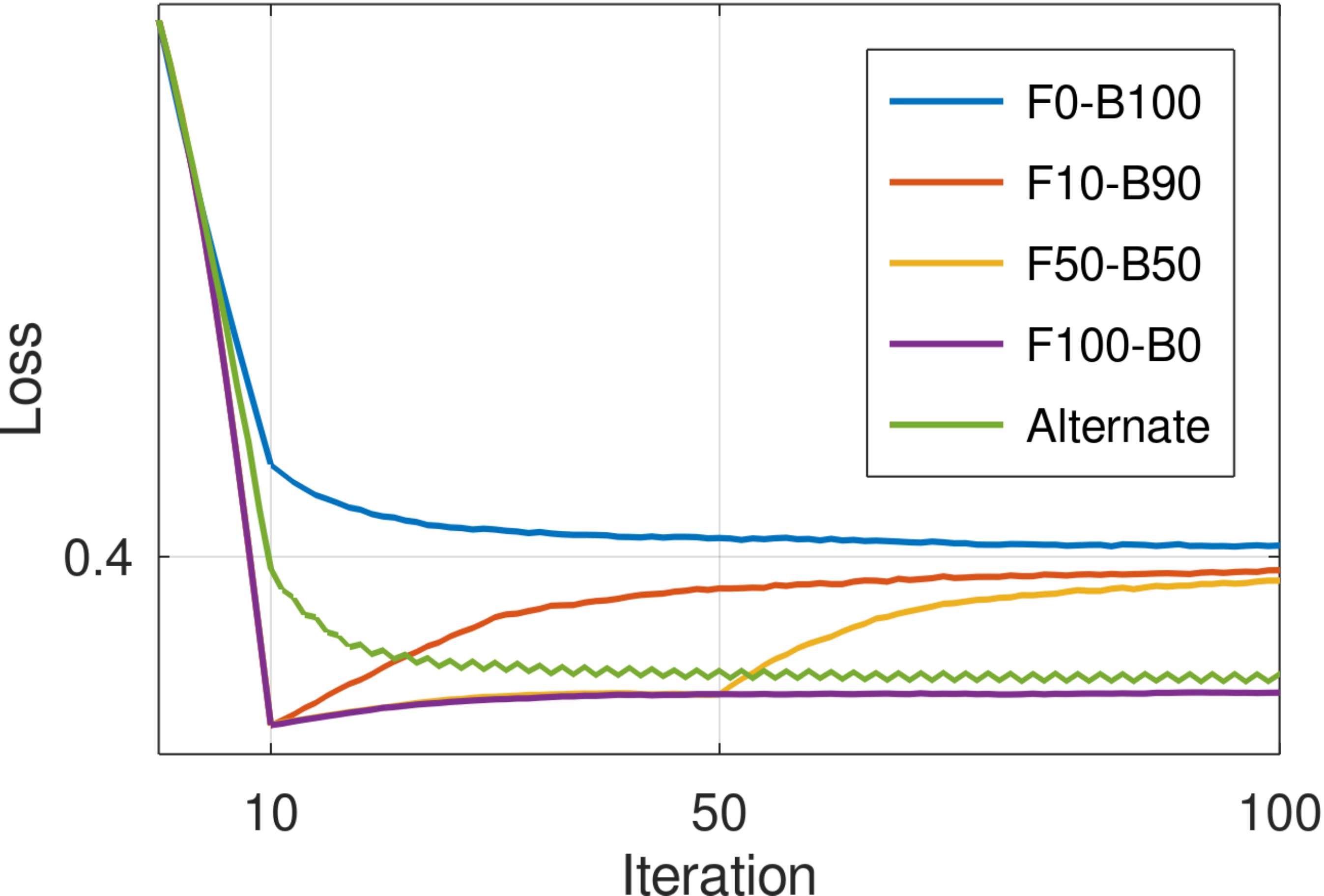}
    \caption{Convergence of cascade learning. Forward-only (F100) reduces loss rapidly but overfits. Backward-only (B100) generalizes better but is slower. Cascade strategies (F10-B90, F50-B50) balance convergence and robustness.}
    \label{fig_loss}
\end{figure}

While each mode has trade-offs, their combination enables a better optimization–robustness balance. The following lemmas formalize this dynamic. Empirical results in \cref{fig_loss} and \cref{tab_2epochs} confirm that cascade learning achieves the best trade-off between fast convergence and transfer robustness.

\begin{lemma}
\label{lemma_cascade_speed}
Forward learning reduces loss more effectively than backward learning:
\begin{equation}
    \mathcal{L}(x) - \mathcal{L}(x_K^{\mathrm{F}}) 
    > 
    \mathcal{L}(x) - \mathcal{L}(x_K^{\mathrm{B}}).
\end{equation}
\end{lemma}

\noindent
This holds because forward learning follows direct gradient descent, while backward learning reverses accumulated ascent, which deviates from the optimal descent path.

\revise{
\begin{lemma}
\label{lemma_cascade_transfer}
Assume that the backward gradient components exhibit non-negligible average alignment with the shared vulnerable subspace $\mathcal{S}(x)$, i.e., their projections onto $\mathcal{S}(x)$ do not vanish in expectation. Then backward learning produces updates that place more energy in the shared vulnerable subspace than forward learning. Formally,
\begin{equation}
    \bigl\| \Pi_{\mathcal{S}} \Delta^{\mathrm{B}} \bigr\|_2^2 
    >
    \bigl\| \Pi_{\mathcal{S}} \Delta^{\mathrm{F}} \bigr\|_2^2 ,
    \label{eq_lemma3}
\end{equation}
where $\Pi_{\mathcal{S}}$ denotes the orthogonal projection onto the shared vulnerable subspace $\mathcal{S}(x)$, and $\Delta^{\mathrm{*}}$ denotes the final perturbation generated by the corresponding update rule ($*$ being forward or backward). The projected norm $\|\Pi_{\mathcal{S}}\Delta^{*}\|_2^2$ measures how much of the perturbation lies in $\mathcal{S}(x)$, \ie the transferable component from the backbone model $f_{\mathrm{back}}$ to the target model $f_{\mathrm{tgt}}$.
\end{lemma}}

\noindent
Forward learning enforces “being right” under \( f_{\mathrm{back}} \), resulting in sharp, class-specific updates that transfer only when \( f_{\mathrm{back}} \approx f_{\mathrm{tgt}} \). In contrast, backward learning enforces “not being wrong,” producing more diffuse updates that broadly avoid errors, leading to more transferable robustness. \revise{The proofs are provided in Appendix~\ref{proof}.}

\subsubsection{\textbf{Gradient Smoothing}}
\label{gradient_smoothing}
To enhance generalization and computational efficiency, we adopt \textit{single-step gradient smoothing}, which averages gradients over $N$ noisy input variants:
\begin{equation}
  \begin{aligned}
      \mathrm{SmoothGrad}(\nabla_x \mathcal{L}) = \frac{1}{N} \sum_{i=1}^N \nabla_x \mathcal{L}(f_\mathrm{back}(x + \xi_i), \hat{y}),
  \end{aligned}
  \label{eq_smartgrad}
\end{equation}
where $\xi_i \sim \mathcal{U}(-\epsilon, \epsilon)$ and $\epsilon$ denotes the $\ell_\infty$ noise bound. This approach reduces sensitivity to high-frequency noise and sharp curvature, yielding smoother, lower-variance gradients that better align with robust directions. Unlike iterative gradient descent, it enables parallel computation and improves generalization \cite{athalye2018synthesizing} (see \cref{tab_ablation_smoothgrad}).

\section{Adaptive Diagnostic: Preemptive Reversion}
\label{preemptivereversion}
\begin{figure}[!t]
    \centering
    \includegraphics[width=\linewidth]{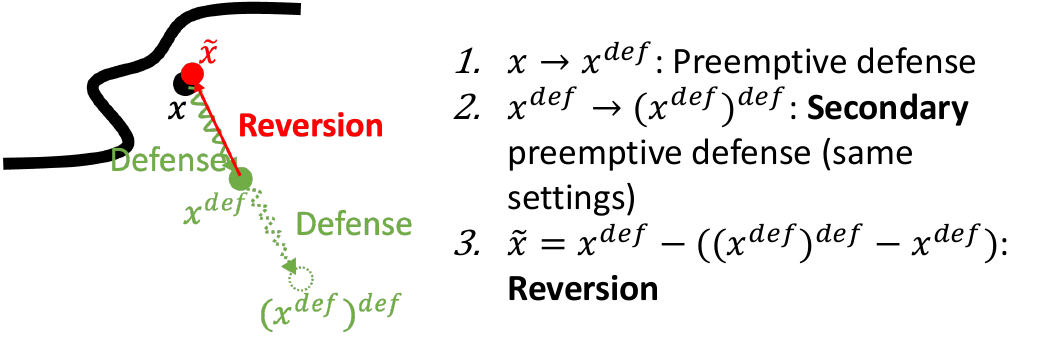}
    \caption{Illustration of the proposed Preemptive Reversion algorithm, which reverses the effect of a secondary preemptive defense with identical settings.}
    \label{fig_ourreversion}
\end{figure}

We introduce Preemptive Reversion (see \cref{fig_ourreversion}), the first diagnostic to effectively remove preemptive signals when the defender’s gradients are fully exposed. JPEG and diffusion purifiers simply distort inputs without neutralizing the protective perturbations (\cref{fig_reversion}). Only our white-box reversion (for stress test only since unrealistic assumption in creator-side workflows) achieves partial cancellation, leaving MSPD +2.2\% more robust than no defense. Our reversion also applies to other preemptive defenses (see \cref{tab_reversion}). 

\subsection{Threat Model}
We define adaptive threat settings as follows:
\begin{itemize}
    \item \textbf{White-Adaptive:} The attacker has full access to the defense pipeline, including the backbone’s architecture and gradients, enabling exact gradient-based adaptation.
    \item \textbf{Gray-Adaptive (Near-White-Adaptive):} The attacker knows the backbone architecture and can train a surrogate on the same dataset, but does not have access to the actual backbone’s weights or gradients. This setting, referred to as the near-white assumption, represents the strongest non-white adaptive threat, as attacks must rely on surrogate models.
    \item \textbf{Black-Adaptive:} The attacker has no information about the backbone and relies on transferability from independently trained surrogates.
\end{itemize}

Gray- and black-adaptive scenarios reflect practical deployments, where the backbone remains hidden. Thus, attackers must rely on approximations. Diffusion-based purifiers and JPEG operate under the black-adaptive assumption.

\subsection{Reversion Attack Algorithm}

We begin with the theoretical insight motivating our reversion design. Lemma~\ref{lemma_reversion} establishes that applying the preemptive defense twice approximates its inverse in white-box settings. 
\begin{lemma}
\label{lemma_reversion}
Let \( \mathrm{Def}(x) = \mathrm{Def}(x, f_{\mathrm{back}}) \) denote a preemptive defense with random initialization applied using backbone model \( f_{\mathrm{back}} \). Define the first defended image as \( x^{\mathrm{def}} = \mathrm{Def}(x) \), and reapplying this defense using a surrogate backbone \( f_{\mathrm{sur}} \) as \( (x^{\mathrm{def}})^{\mathrm{def}}_\mathrm{sur} = \mathrm{Def}_\mathrm{sur}(x^{\mathrm{def}}) = \mathrm{Def}(x^{\mathrm{def}}, f_{\mathrm{sur}}) \).

In the white-box setting (\( \mathrm{Def} = \mathrm{Def}_\mathrm{sur}\)), such reapplying approximately reconstructs the first perturbations:
\begin{equation}
    (x^{\mathrm{def}})^{\mathrm{def}} - x^{\mathrm{def}} \approx x^{\mathrm{def}} - x.
\end{equation}
In gray- or black-box settings (\( \mathrm{Def} \ne \mathrm{Def}_\mathrm{sur} \)), this approximation deteriorates due to gradient mismatch.
\end{lemma}
\noindent 

Thus, the attacker can reverse the preemptive defense by subtracting the second perturbation:
\begin{equation}
    \tilde{x} = x^{\mathrm{def}} - \left((x^{\mathrm{def}})^{\mathrm{def}}_\mathrm{sur} - x^{\mathrm{def}}\right) = 2x^{\mathrm{def}} - (x^{\mathrm{def}})^{\mathrm{def}}_\mathrm{sur}.
    \label{eq_same_delta}
\end{equation}
In the white-box case, \cref{eq_same_delta} gives $\tilde{x} \approx x$, indicating that the clean input is nearly recovered. As shown in \cref{fig_reversion}, only white-box reversion mostly cancels the defense, restoring perceptual similarity and vulnerability, whereas gray/black-box or purification variants fail due to gradient misalignment, showing the need for full backbone access. \revise{The proof is provided in Appendix~\ref{proof}.}

\begin{table}[!t]
    \scriptsize
    \centering
    \begin{threeparttable}
        \caption{Experimental settings.}
        \label{tab_settings}
        \setlength{\tabcolsep}{0.6mm}{\begin{tabular}{cc}
            \toprule
            Type&Settings\\
            \midrule
            Baseline Preemptive Defense&$\ell_\infty$-norm $\epsilon$ = $\frac{8}{255}^a/\frac{4}{255}^b$, Epoch=100, $\alpha$=0.1\\
            \midrule
            Adversarial Attack&\begin{tabular}{c}$\ell_\infty$-norm $\epsilon$ = $\frac{8}{255}^a/\frac{4}{255}^b$, $\ell_2$-norm $\epsilon$ = $0.5^a/2^b$\\Maximum Queries = 10,000, EOT=20\end{tabular}\\
            \midrule
            Purification&\begin{tabular}{c}JPEG Quality Factor = 50\%; DiffPure Timestep = 0.1\end{tabular}\\
            \midrule
            \textbf{Our Defense and Reversion}&\begin{tabular}{c}$\ell_\infty$-norm $\epsilon$ = $\frac{8}{255}^a/\frac{4}{255}^b$, Epoch=2, $\alpha$=1, N=20\end{tabular}\\
            \bottomrule
        \end{tabular}}
        \begin{tablenotes}
            \item $^a$ for CIFAR-10; $^b$ for ImageNet.
        \end{tablenotes}
    \end{threeparttable}
\end{table}

\section{Experiment Settings}
\label{settings}
The experimental settings are detailed in \cref{tab_settings}. All experiments were performed on a single NVIDIA A100 GPU. The attack budget $\epsilon$ is computed with respect to the inputs (either the original or the defended images), rather than only the original inputs. Consequently, when attacking defended images, the $L_\infty$ distance from the original inputs can reach up to $2 \times \epsilon$, making the evaluation more challenging from the defense perspective. Note that the defense budget $\epsilon$ does not need to be aligned with the attack budget $\epsilon$, as further discussed in \cref{ablation_eps}.

\textbf{Datasets and Models}
We evaluate on CIFAR-10 (low-resolution, $32 \times 32$)~\cite{krizhevsky2009learning} and ImageNet (high-resolution, $224 \times 224$)~\cite{nips2017}, the \emph{canonical pair} spanning frequency and scale extremes in robustness studies (\eg RobustBench~\cite{croce2021robustbench}). For CIFAR-10, we use a ResNeXt29-32x4d classifier~\cite{hendrycks2019augmix} and a PreActResNet-18 backbone~\cite{gowal2021improving}. For ImageNet, we pair a DeiT-Base classifier~\cite{tian2022deeper} with a ViT-S backbone~\cite{singh2023revisiting}. Both backbones are \emph{public, lightweight} (11.2M and 22M parameters), incurring no extra training cost (see \cref{tab_ablation_backbone} for backbone selection). To assess generalization, we test on \textit{11 unseen target models}, detailed in \cref{tab_defense} and \cref{tab_transferability}.

\textbf{Baselines}
We use Bi-Level~\cite{moon2022preemptive} as the \textit{backward-only} baseline, leveraging its public implementation. This SOTA method couples the classifier and backbone via a fine-tuned adversarially trained model and uses backward learning with iterative gradient descent, without smoothing. For the \textit{forward-only} baseline, we adapt UnAdv~\cite{salman2021unadversarial}, originally for physical corruption robustness, to our digital threat setting. For completeness, we also compare the proposed preemptive defenses against the purification-based test-time method DiffPure~\cite{nie2022diffusion}. However, since our focus is not on purification-based defenses, this experiment serves only as a reference comparison rather than a direct competitor to our approach. The No Preemptive (No Pre.) baseline corresponds to adversarial training.

\textbf{Attacks}
\revise{We evaluate robustness against \textit{eight diverse attacks}, including one adaptive (\cref{tab_attack})~\cite{carlini2017adversarial,madry2018towards,croce2020reliable,chen2020rays,andriushchenko2020square,yin2023generalizable,lin2024boosting,athalye2018obfuscated}, two purification-based reversions (\cref{fig_reversion})~\cite{nie2022diffusion}, and our proposed adaptive reversion. These span white-, black-box, and adaptive threat models; $\ell_2$ and $\ell_\infty$ norms; and attack types: gradient-based (white-box) and gradient-free (black-box). For AutoAttack~\cite{croce2020reliable}, both standard and randomized variants are evaluated.} Unless stated otherwise, when evaluating attacks on defended images we constrain perturbations relative to the defended input $x^{\mathrm{def}}$ with budget $\epsilon$. The defense budget is disjoint from the attacker’s budget. When attacks are instead computed from the original input $x$, the maximum $L_\infty$ distance between $x^{\mathrm{def}}$ and $x$ can reach $2\epsilon$.

\textbf{Evaluation Metrics.}
\revise{We report clean and robust accuracy (higher is better) to measure defense performance. Visual fidelity is evaluated using $L_\infty$ and $L_2$ distances, SSIM~\cite{wang2004image}, and LPIPS~\cite{zhang2018unreasonable} relative to the original inputs. Higher SSIM and lower distances and LPIPS indicate better perceptual quality. These quantitative metrics are further supported by a user study\footnote{\label{foot_userstudy}https://github.com/azrealwang/MSPD/blob/main/UserStudy.png} involving 20 participants and 200 images (4,000 total evaluations).}

\begin{table}[!t]
    \scriptsize
    \centering
    \begin{threeparttable}
        \caption{Preemptive defenses against AutoAttack.}
        \label{tab_defense}
        \setlength{\tabcolsep}{1.9mm}{\begin{tabular}{ccccccc}
            \toprule
            Dataset&Defense&Time (s)&Clean&Std$\ell_\infty$&Std$\ell_2$&Rand$\ell_\infty$\\
            \midrule
            \multirow{4}{*}{\begin{tabular}{c}CIFAR-10\\($32 \times 32$)\end{tabular}}&No Pre. &N/A&93.9&74.0&70.4&74.6\\
            &UnAdv&38.36&93.1&74.0&71.2&74.3\\
            &Bi-Level&33.92&92.4&82.2&78.4&82.2\\
            &\textbf{Ours}&\textbf{\ 0.02}&\textbf{95.9}&\textbf{86.3}&\textbf{83.1}&\textbf{86.4}\\
            \midrule
            \multirow{4}{*}{\begin{tabular}{c}ImageNet\\($224 \times 224$)\end{tabular}}&No Pre.&N/A&93.0&67.0&59.3&67.5\\
            &UnAdv&55.13&93.4&70.1&61.9&70.9\\
            &Bi-Level&37.84&93.6&71.6&62.9&72.4\\
            &\textbf{Ours}&\textbf{\ 0.26}&\textbf{94.9}&\textbf{78.2}&\textbf{68.2}&\textbf{79.1}\\
            \bottomrule
        \end{tabular}}
        \begin{tablenotes}
            \item Our defense significantly reduces image processing time while improving both clean and robust accuracy. Baselines are evaluated under their original 100-epoch training settings to ensure optimal robustness. We use WideResNet-94-16~\cite{bartoldson2024adversarial} as the target model for CIFAR-10 and Swin-L~\cite{liu2024comprehensive} for ImageNet. \textbf{Bold} indicates the best performance.
        \end{tablenotes}
    \end{threeparttable}
\end{table}

\begin{table}[!t]
    \scriptsize
    \centering
    \begin{threeparttable}
        \caption{Adversarial defenses against unseen attacks.}
        \label{tab_attack}
        \setlength{\tabcolsep}{0.6mm}{\begin{tabular}{ccccccc}
            \toprule
            Attack&Type&No Pre.&DiffPure$^1$&UnAdv&Bi-Level&\textbf{Ours}\\
            \midrule
            CW&Subtle White-Box&80.8&85.3&82.6&85.6&\textbf{90.0}\\
            PGD&Efficient White-Box&77.1&76.5&77.8&84.5&\textbf{88.1}\\
            AutoAttack&Strongest White-Box&74.0&79.4&74.0&82.2&\textbf{86.3}\\
            \midrule
            \revise{RayS}&\revise{Query-based Black-Box}&86.7&83.0&86.7&87.9&\textbf{93.9}\\
            Square&Query-based Black-Box&77.2&80.5&77.8&83.8&\textbf{87.8}\\
            MCG&Query-based Black-Box&76.9&80.1&77.4&83.3&\textbf{87.7}\\
            DeCoWA&Transferable Black-Box&87.8&81.1&88.7&89.6&\textbf{93.0}\\
            \midrule
            BPDA+EOT20&Adaptive White-Box&76.7&74.5&77.4&84.1&\textbf{87.7}\\
            \bottomrule
        \end{tabular}}
        \begin{tablenotes}
            \item $^1$ A test-time defense with costly sampling, in contrast to preemptive, one-shot design. All attacks use an $\ell_\infty$ perturbation budget of $\epsilon = 8/255$. The target model is WideResNet-94-16~\cite{bartoldson2024adversarial}. \textbf{Bold} indicates the best performance.
        \end{tablenotes}
    \end{threeparttable}
\end{table}

\begin{table}[!t]
    \scriptsize
    \centering
    \begin{threeparttable}
        \caption{Preemptive defenses for ten unseen models.}
        \label{tab_transferability}
        \setlength{\tabcolsep}{1.5mm}{\begin{tabular}{ccccccc}
            \toprule
            Model&Metric&No Pre.&DiffPure$^1$&UnAdv&Bi-Level&\textbf{Ours}\\
            \midrule
            \multirow{2}{*}{WideResNet-94-16}&Clean&93.9&91.9&93.1&92.4&\textbf{95.9}\\
            &Robust&74.0&79.4&74.0&82.2&\textbf{86.3}\\
            \multirow{2}{*}{RaWideResNet-70-16}&Clean&93.5&91.2&92.8&92.0&\textbf{95.3}\\
            &Robust&71.5&77.9&72.0&81.0&\textbf{86.0}\\
            \multirow{2}{*}{WideResNet-70-16}&Clean&93.3&91.2&92.3&91.8&\textbf{95.3}\\
            &Robust&71.2&76.2&71.6&81.0&\textbf{85.8}\\
            \multirow{2}{*}{WideResNet-34-10}&Clean&91.5&89.3&90.9&90.4&\textbf{95.0}\\
            &Robust&62.1&73.7&64.1&76.0&\textbf{80.9}\\
            \multirow{2}{*}{WideResNet-28-10}&Clean&93.4&91.4&92.5&91.9&\textbf{95.7}\\
            &Robust&64.8&74.7&65.9&77.4&\textbf{82.9}\\
            \multirow{2}{*}{PreActResNet-18$^2$}&Clean&84.1&84.0&86.1&87.3&\textbf{91.6}\\
            &Robust&42.4&\textbf{62.6}&43.5&55.1&61.5\\
            \multirow{2}{*}{ResNeSt-152}&Clean&86.9&86.8&87.8&87.8&\textbf{92.5}\\
            &Robust&62.7&69.9&63.9&73.1&\textbf{78.5}\\
            \multirow{2}{*}{ResNet-50}&Clean&86.6&85.5&88.3&88.6&\textbf{93.2}\\
            &Robust&52.0&65.3&54.7&67.7&\textbf{72.0}\\
            \multirow{2}{*}{ResNet-18}&Clean&86.6&84.6&87.7&88.4&\textbf{92.6}\\
            &Robust&51.4&67.2&54.1&67.1&\textbf{72.9}\\
            \multirow{2}{*}{XCiT-L12}&Clean&92.8&91.2&91.8&91.5&\textbf{95.3}\\
            &Robust&57.9&70.3&59.3&70.4&\textbf{76.2}\\
            \multirow{2}{*}{\textit{Average}}&Clean&90.3&88.7&90.3&90.2&\textbf{94.2}\\
            &Robust&61.0&71.7&62.3&73.1&\textbf{78.3}\\
            \bottomrule
        \end{tabular}}
        \begin{tablenotes}
            \item $^1$ A test-time defense with costly sampling, in contrast to preemptive, one-shot design. $^2$ Same architecture as the backbone, but with different weights. Baselines often trade clean accuracy for robustness and still underperform. All attacks are AutoAttack (Std) under an $\ell_\infty$ constraint with $\epsilon = 8/255$. \textbf{Bold} denotes the best result.
        \end{tablenotes}
    \end{threeparttable}
\end{table}

\begin{table}[!t]
    \scriptsize
    \centering
    \begin{threeparttable}
        \caption{Forward/backward-only \vs naive cascade \vs ours.}
        \label{tab_2epochs}
        \setlength{\tabcolsep}{2.6mm}{\begin{tabular}{ccccc}
            \toprule
            Algorithm&Defense&Time (s)&Clean&Robust\\
            \midrule
            \multirow{2}{*}{Forward}&UnAdv (100 Epochs)&38.36&93.1&74.0\\
            &UnAdv (2 Epochs)&\ 0.55&93.1&77.1\\
            \midrule
            \multirow{2}{*}{Backward}&Bi-Level (100 Epochs)&33.92&92.4&82.2\\
            &Bi-Level (2 Epochs)&\ 0.56&92.8&80.5\\
            \midrule
            \multirow{3}{*}{Cascade}&UnAdv + Bi-Level (2 Epochs)&\ 0.49&92.8&80.9\\
            &\textbf{Ours} (2 Epochs)&\textbf{\ 0.02}&\textbf{95.9}&\textbf{86.3}\\
            &\textbf{Ours} (4 Epochs)&\ 0.04&\textbf{95.9}&86.2\\
            \bottomrule
        \end{tabular}}
        \begin{tablenotes}
            \item With limited training epochs, the SOTA baseline (Bi-Level) is less effective. Incorporating forward learning via cascade offers slight improvement but still lags behind our method due to the lack of a decoupled pipeline and gradient smoothing. All attacks are AutoAttack (Std) under an $\ell_\infty$ constraint with $\epsilon = 8/255$. The target model is WideResNet-94-16~\cite{bartoldson2024adversarial}. \textbf{Bold} indicates the best performance. 
        \end{tablenotes}
    \end{threeparttable}
\end{table}

\section{Performance}
\label{performance}

\subsection{Preemptive Defense Performance}
\label{performance_defense}

We evaluate preemptive defenses against eight unseen (including adaptive) attacks across two datasets and 11 unseen target models. As shown in \cref{tab_defense,tab_attack,tab_transferability,tab_2epochs}, our method consistently outperforms no defense and prior preemptive baselines. Compared to no defense, it improves clean accuracy by 1.8--7.5\% (avg. +3.8\%) and robust accuracy by 5.2--21.5\% (avg. +13.4\%). Over the prior SOTA (Bi-Level), it achieves +3.7\% clean and +5.0\% robust accuracy on average, while being 28--1696$\times$ faster. All baselines use their original 100-epoch budgets. Under equal limited training (2 epochs; \cref{tab_2epochs}), results validate \cref{lemma_cascade_speed} and \cref{lemma_cascade_transfer}. Forward-only converges quickly but overfits; backward-only generalizes better but converges slowly. Cascade improves over both, yet without our decoupled pipeline and gradient smoothing remains inferior. Additional epochs provide marginal gains at higher cost, supporting our ``minimal and sufficient'' design. Finally, \cref{tab_a5} evaluates the generative preemptive baseline A$^5$~\cite{frosio2023best} under our $\ell_\infty$ budget ($\epsilon = 8/255$). Its clear performance gap highlights the difficulty of adapting physical-world defenses to fine-grained digital threat settings.

\begin{table}[!t]
    \scriptsize
    \centering
    \begin{threeparttable}
        \caption{Performance of physical generator-based preemptive defense A$^5$~\cite{frosio2023best} against AutoAttack under small visual perturbation.}
        \label{tab_a5}
        \setlength{\tabcolsep}{2.2mm}{\begin{tabular}{ccccccc}
            \toprule
            Dataset&Defense&Time (s)&Clean&Std$L_\infty$&Std$L_2$&Rand$L_\infty$\\
            \midrule
            \multirow{3}{*}{CIFAR-10}&No Pre. &N/A&93.9&74.0&70.4&74.6\\
            &A$^5$&\textbf{0.01}&93.3&73.7&70.2&74.4\\
            &\textbf{Ours}&0.02&\textbf{95.9}&\textbf{86.3}&\textbf{83.1}&\textbf{86.4}\\
            \bottomrule
        \end{tabular}}
        \begin{tablenotes}
            \item A$^5$~\cite{frosio2023best} shows limited effectiveness under our setting, which uses a small visual perturbation budget ($L_\infty$-norm, $\epsilon = 8/255$). The target model is WideResNet-94-16~\cite{bartoldson2024adversarial} on CIFAR-10. \textbf{Bold} denotes the best performance.
        \end{tablenotes}
    \end{threeparttable}
\end{table}

\begin{figure}[!t]
    \centering
    \includegraphics[width=\linewidth]{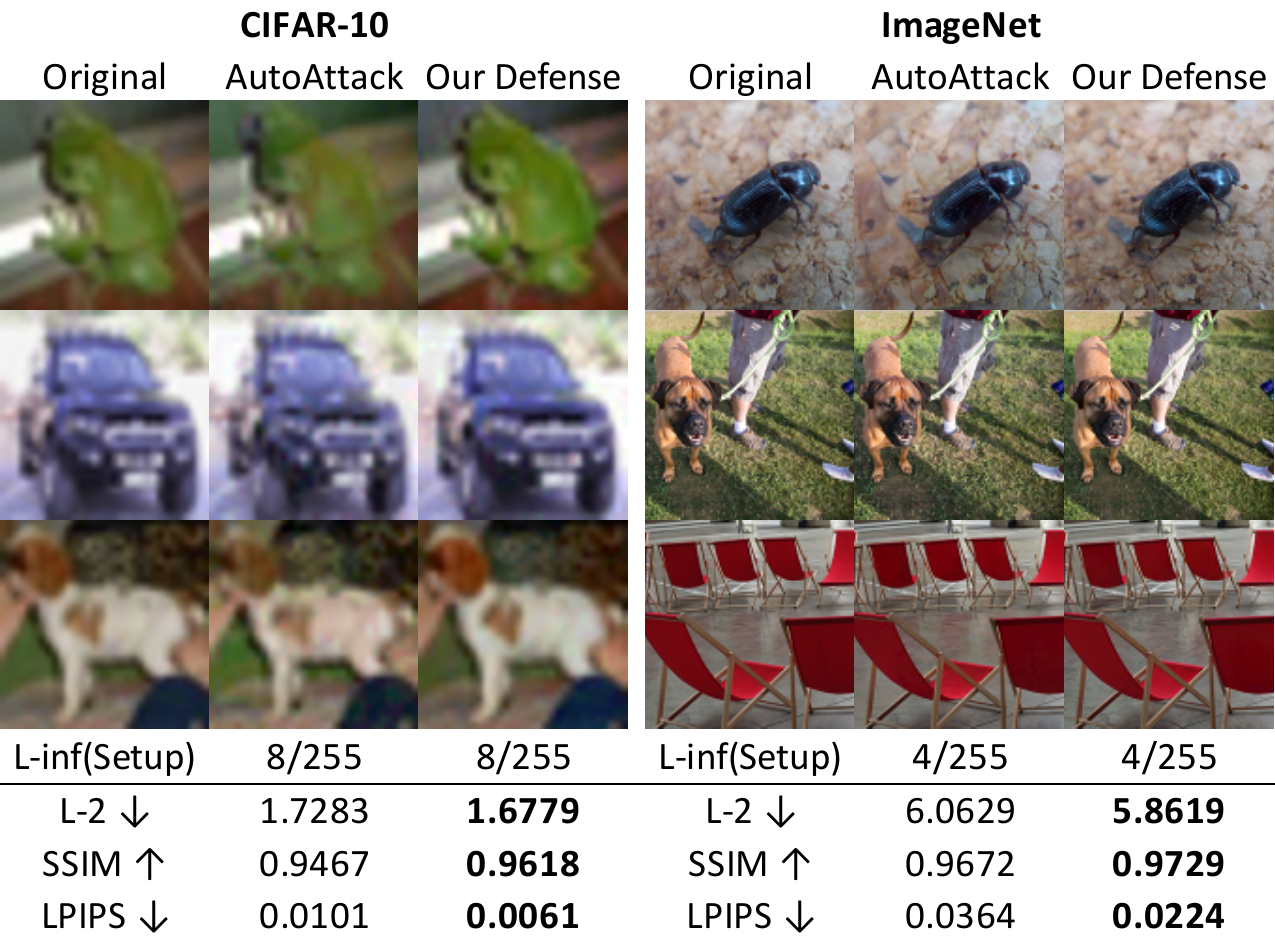}
    \caption{\revise{Dataset-wide average visual quality of protected images. Our defense achieves higher SSIM and lower LPIPS than adversarial examples under the same $\ell_\infty$ constraint. Since such adversarial examples are widely considered visually acceptable~\cite{goodfellow2014explaining,madry2018towards,croce2020reliable}, this indicates our defense meets or exceeds established perceptual quality standards.}}
    \label{fig_visualquality}
\end{figure}

\begin{reviseG}

\subsection{Qualitative Analysis}
\label{qualitative}

For practical deployment, protected images should remain the visual quality from the originals. As shown in \cref{fig_visualquality}, our method consistently achieves higher perceptual quality (higher SSIM, lower LPIPS) than adversarial examples under the same $\ell_\infty$ constraint. Since such adversarial examples are generally regarded as visually acceptable~\cite{goodfellow2014explaining,madry2018towards,croce2020reliable}, this suggests our defense preserves perceptual quality. This advantage stems from MCGS using a robust backbone, which produces perturbations aligned with semantic image features (see \cref{fig_gradient_similarity}), concentrating on edges and complex textures rather than introducing diffuse high-frequency noise, as commonly observed in PGD~\cite{wang2025greedy}.

\textbf{User Study.}
We conducted a user study with 20 participants and 200 images (4,000 participant–image evaluations)\footref{foot_userstudy}, comparing 100 MCGS-protected images with 100 PGD examples. Images were evaluated under five conditions: raw output and four post-processing transformations (sharpening, high brightness, high contrast, compression). Participants identified modified images whose visual quality they judged worse than the original. As summarized in \cref{tab_userstudy}, MCGS images are rarely judged as degraded, even under sharpening, which explicitly amplifies perturbations. Overall, MCGS substantially outperforms PGD in preserving perceptual quality.

\begin{table}[!t]
    \scriptsize
    \centering
    \begin{threeparttable}
        \caption{\revise{User-study results: rate at which images are judged to have degraded visual quality after defense and post-processing. Lower is better.}}
        \label{tab_userstudy}
        \setlength{\tabcolsep}{1.5mm}{\begin{tabular}{l|ccccc|c}
            \toprule
            Algorithm&Raw&Sharpen&\makecell{High\\Brightness}&\makecell{High\\Contrast}&Compression&All\\
            \midrule
            PGD&3/20&\underline{6/20}&4/20&4/20&3/20&20/100\\
            \textbf{MCGS (Ours)}&\textbf{1/20}&\underline{\textbf{2/20}}&\textbf{1/20}&\textbf{1/20}&\textbf{1/20}&\textbf{6/100}\\
            \bottomrule
        \end{tabular}}
        \begin{tablenotes}
            \item \revise{All defenses use $\ell_\infty$ constraint with $\epsilon=8/255$. \textbf{Bold} indicates better perceptual quality. \underline{Underline} denotes the strongest enhancement (sharpening).}
        \end{tablenotes}
    \end{threeparttable}
\end{table}

\textbf{Failure Cases.}
Rare failures occur when the original image contains overly simple textures (e.g., smooth backgrounds), where small perturbations become perceptible, especially after sharpening (\cref{fig_quality_degrade}). Even in these cases, MCGS introduces markedly less visible noise than PGD.

\begin{figure}[!t]
    \centering
    \includegraphics[width=\linewidth]{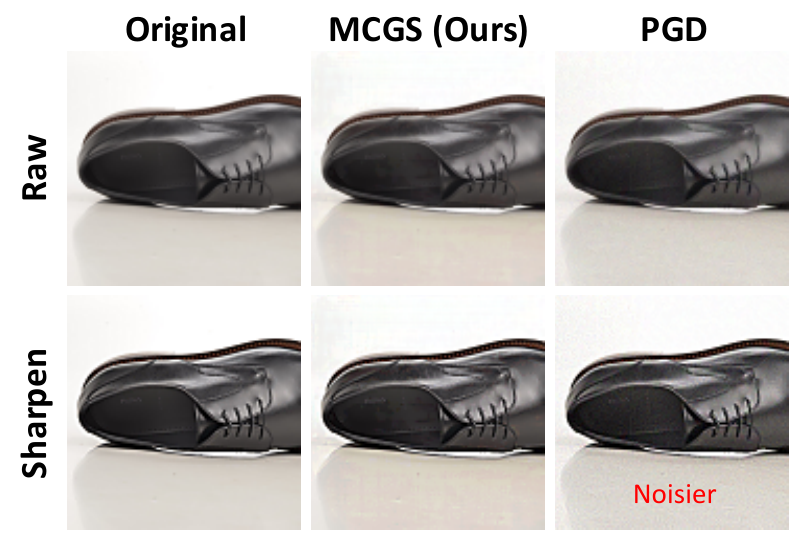}
    \caption{\revise{Representative failure case where quality degradation is perceived due to overly simple textures in the original image. MCGS introduces substantially less visible noise than PGD.}}
    \label{fig_quality_degrade}
\end{figure}

\end{reviseG}

\subsection{Adaptive Diagnostic}
\label{performance_reversion}
Our defense improves both clean and robust accuracy on unseen target models. To evaluate resilience against compression and adaptive threats, we define successful reversion as one that \textit{simultaneously}: \one reduces accuracy to the undefended level, and \two increases SSIM while reducing LPIPS toward the original. This ensures degradation stems from true reversal, not distortion. We use this protocol to validate \cref{lemma_reversion}.

\begin{figure}[!t]
    \centering
    \includegraphics[width=\linewidth]{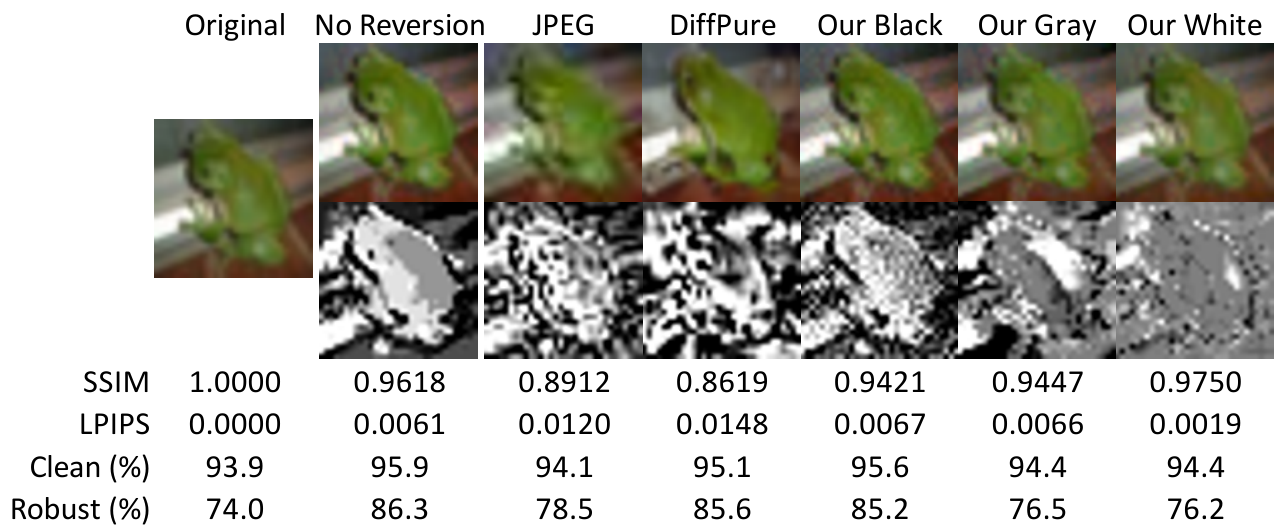}
    \caption{Illustration of reversion attacks. Under the strongest but unrealistic white-box assumption, our defense still achieves +2.2\% higher robustness than the no-defense baseline. Under all weaker (non–white-box) conditions, the defense cannot be reversed, and attempted reversion only distorts the images.}
    \label{fig_reversion}
\end{figure}

As shown in \cref{fig_reversion}, only our adaptive reversion achieves partial success, exclusively in the white-box setting with backbone gradient access, an unrealistic assumption. Even then, robustness remains +2.2\% above the undefended case. In contrast, gray-box adaptive attacks yield limited reversal, and purification-based methods distort images without canceling the defense. These results support \cref{lemma_reversion}.

\begin{table}[!t]
    \scriptsize
    \centering
    \begin{threeparttable}
        \caption{Generalization of the proposed Preemptive Reversion on various preemptive defenses.}
        \label{tab_reversion}
        \setlength{\tabcolsep}{1.9mm}{\begin{tabular}{cccccc}
            \toprule
            Defense&Reversion&SSIM&LPIPS&Clean (\%)&Robust (\%)\\
            \midrule
            No Pre.&N/A&1.0000&0.0000&93.9&74.0\\
            \midrule
            \multirow{2}{*}{UnAdv (2 Epochs)}&No&0.9570&0.0066&93.1&77.1\\
            &White&0.9632&0.0028&93.6&75.1\\
            \midrule
            \multirow{2}{*}{Bi-Level (2 Epochs)}&No&0.9659&0.0046&92.8&80.5\\
            &White&0.9563&0.0041&93.6&73.6\\
            \midrule
            \multirow{2}{*}{\textbf{Ours} (2 Epochs)}&No&0.9618&0.0061&95.9&86.3\\
            &White&0.9750&0.0019&94.4&76.2\\
            \bottomrule
        \end{tabular}}
        \begin{tablenotes}
            \item All reversed images move closer to those of the no-defense baseline, demonstrating effective reversion attacks. 
            \item All attacks are conducted using AutoAttack (Standard) under an \( \ell_\infty \) constraint with \( \epsilon = 8/255 \). The target model is WideResNet-94-16~\cite{bartoldson2024adversarial}.
        \end{tablenotes}
    \end{threeparttable}
\end{table}

\textbf{Generalization of Preemptive Reversion to Other Defenses}
As formally justified in \cref{lemma_reversion}, the proposed preemptive reversion attack is not specifically tailored to our MSPD method, but instead generalizes to other gradient-based preemptive defenses, including the established baselines UnAdv~\cite{salman2021unadversarial} and Bi-Level~\cite{moon2022preemptive}. As shown in \cref{tab_reversion}, both UnAdv and Bi-Level exhibit similar vulnerabilities under white-box adaptive reversion, akin to those observed for MSPD, bringing all reversed images closer to those of the no-defense baseline. Specifically, the robust accuracy of all defenses decreases after reversion, confirming the vulnerability of gradient-based preemptive methods. While our MSPD slightly improves clean accuracy prior to reversion, the other two baselines (UnAdv and Bi-Level) degrade it. After reversion, this trend reverses: MSPD experiences a drop in clean accuracy, whereas the baselines recover, with accuracy levels approaching those of the original undefended model. In terms of image quality, all defenses introduce perceptual distortion—evidenced by reduced SSIM and increased LPIPS, while reversion generally restores quality. A notable exception is Bi-Level, whose SSIM remains low post-reversion, although LPIPS improves as expected; this discrepancy is left as unclear. These findings demonstrate not only the effectiveness of our reversion attack but also its generalizability as a diagnostic tool, capable of stress-testing a broad class of preemptive defense strategies.

\section{Ablation Studies}
\label{ablation}
We conduct ablation studies under fixed settings: CIFAR-10, $\ell_\infty$-bounded PGD attacks with $\epsilon = 8/255$, and a fixed target model (WideResNet-94-16~\cite{bartoldson2024adversarial}).

\begin{table}[!t]
    \scriptsize
    \centering
    \begin{threeparttable}
        \caption{Backbone gradient alignment and analysis of success–failure cases.}
        \label{tab_ablation_backbone}
        \setlength{\tabcolsep}{0.4mm}{\begin{tabular}{c|c|c|c|c|c|c}
            \toprule
            Backbone&Params&Grad Sim&Time&Clean&Robust&$\mathcal{L}_\mathrm{tgt}^1$\\
            \midrule
            No Pre. Defense&N/A&N/A&N/A&92.8&60.2&0.39/0.30/0.52\\
            \midrule
            \revise{ResNet-56 (Nonrobust)}&\ \ 0.9M&0.0052&0.02&92.8&61.0&0.28/0.22/0.47\\
            WideResNet-28-10 (Nonrobust)&\ 36.5M&0.0124&0.04&92.6&61.2&0.38/0.31/0.49\\
            WideResNet-28-10&\ 36.5M&0.2093&0.04&95.0&65.6&0.36/0.28/0.48\\
            WideResNet-94-16&356.0M&0.4838&0.23&95.2&77.3&0.28/0.23/0.45\\
            PreActResNet-18 (\textbf{Ours})&\ 11.2M&0.4904&0.02&95.3&78.7&0.27/0.22/0.47\\
            \bottomrule
        \end{tabular}}
        \begin{tablenotes}
            \item \revise{$^1$ Average loss (cross-entropy) is reported for all/success/failure cases over 100 images.} Gradient similarity is the mean cosine similarity between backbone and target gradients. Lighter backbones yield faster inference, and higher gradient similarity correlates with lower loss on target models and better clean/robust accuracy. The target model is fixed as XCiT-L12~\cite{debenedetti2023light} (Transformer).
        \end{tablenotes}
    \end{threeparttable}
\end{table}

\begin{reviseG}
\subsection{Backbone-Target Gradient Alignment vs. Robustness}
\label{ablation_backbone}
To validate \cref{lemma_backbone} on the importance of backbone–target gradient alignment, we evaluate five backbones against the same unseen transformer target. As shown in \cref{tab_ablation_backbone}, lighter backbones (fewer parameters) offer lower processing time, while higher gradient similarity (\textit{grad sim}) correlates with better clean and robust accuracy, reflected by lower average loss on the target (all starting from the same clean image, but ending with higher confidence on the ground truth). Lower loss corresponds to fewer failures (high-loss regions) and more successes (low-loss regions). Notably, heavier architectures do not guarantee better generalization, and nonrobust models yield near-zero gradient similarity, failing to improve robustness. These findings empirically support \cref{lemma_backbone}.

\begin{figure}[!t]
    \centering
    \includegraphics[width=\linewidth]{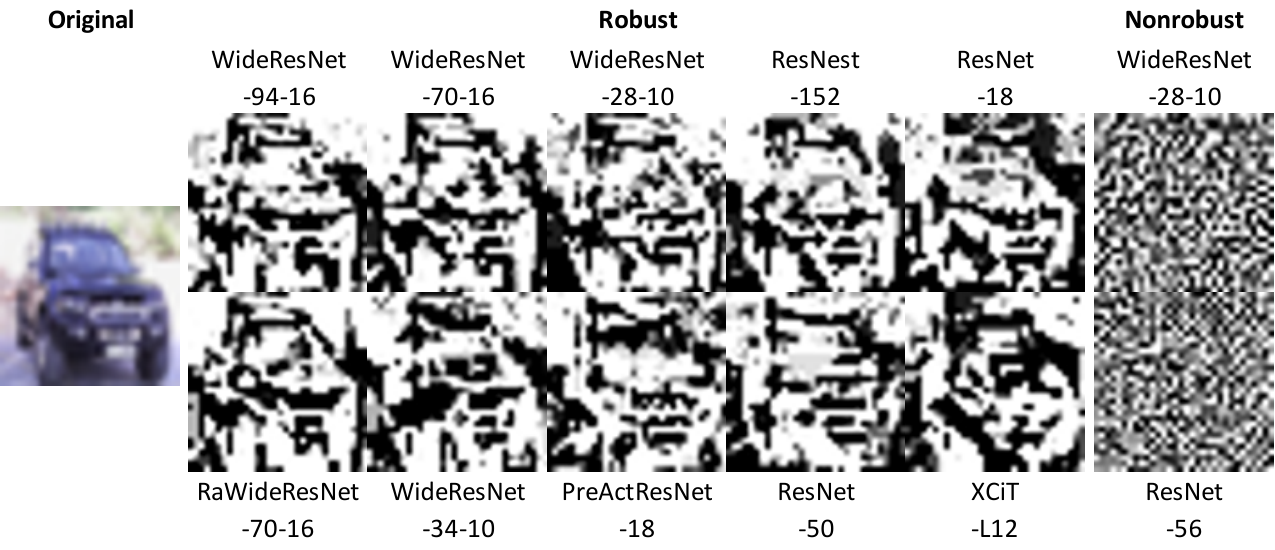}
    \caption{\revise{Illustration of learned perturbations across different models. Robust models exhibit similar car-shaped perturbation patterns, whereas nonrobust models produce unstructured noise. This observation is consistent with the gradient similarity results reported in \cref{tab_ablation_backbone}.}}
    \label{fig_gradient_similarity}
\end{figure}

\cref{fig_gradient_similarity} further supports our claim in \cref{lemma_backbone} that robust models exhibit higher gradient similarity, resulting in visually similar perturbation patterns. This similarity enables proactive perturbations learned from one robust model to transfer effectively to other robust models. In contrast, nonrobust models show substantially weaker inter-model correlations, with especially limited correlation between robust and nonrobust models. Consequently, using nonrobust models as backbones yields little robustness improvement.

\textbf{Success and Failure Analysis.} We compute the cross-entropy loss values separately for successful, failed, and overall cases (lower values indicate more correct classification), and report them in \cref{tab_ablation_backbone}. The results in the last column show a substantial gap (approximately 0.2 in cross-entropy) between success and failure cases. Failures are primarily attributable to model misalignment, as reflected by significantly higher cross-entropy loss compared to the average, whereas successful cases consistently achieve much lower loss values.

\textbf{Limitation.}
In summary, while robust backbones consistently improve defense robustness and lightweight backbones accelerate learning, we do not observe a clear or consistent correlation between backbone size, intrinsic robustness, or architecture and the resulting preemptive defense performance. Consequently, there is currently no principled strategy for selecting an ``optimal'' backbone beyond empirical validation. Understanding this relationship remains an open problem and an important direction for future work.

\end{reviseG}

\subsection{Decoupling Pipeline and Prior Knowledge}
We configure the defense as \textbf{classifier $\neq$ backbone $\neq$ target} to reflect realistic deployment. To assess the role of prior knowledge, we compare variants with access to target gradients or ground-truth labels. As shown in \cref{tab_ablation_pipeline}, decoupling the classifier from the backbone improves robustness with little cost. Additional knowledge further boosts robustness. These empirically corroborate \cref{lemma_backbone} and classifier selection.

\begin{table}[!t]
    \scriptsize
    \centering
    \begin{threeparttable}
        \caption{Decoupling pipeline and more prior knowledge.}
        \label{tab_ablation_pipeline}
        \setlength{\tabcolsep}{1.4mm}{\begin{tabular}{cccc}
            \toprule
            Pipeline&Time (s)&Clean (\%)&Robust (\%)\\
            \midrule
            Classifier$=$Backbone$\neq$Target&0.02&91.0&86.0\\
            Classifier$\neq$Backbone$\neq$Target (Black, \textbf{Ours})&0.02&95.9&88.1\\
            Classifier$\neq$Backbone$=$Target (White)&0.23&95.9&91.0\\
            Ground Truth+Backbone$\neq$Target (GT+Black)&0.02&98.2&89.5\\
            Ground Truth+Backbone$=$Target (GT+White)&0.23&99.0&92.9\\
            \bottomrule
        \end{tabular}}
        \begin{tablenotes}
            \item Our pipeline improves both clean and robust accuracy. Incorporating direct access to the target model or ground-truth labels (impractical in real-world scenarios) can further enhance defense effectiveness.
        \end{tablenotes}
    \end{threeparttable}
\end{table}

\begin{reviseG}
\subsection{Number of Epochs in Cascade Learning}
As shown in \cref{tab_2epochs}, our cascade learning strategy consistently outperforms forward-only, backward-only, and simple forward-backward combination baselines. Here, we clarify the epoch configuration adopted in cascade learning, which motivates the name of our method, \emph{``Minimal but Sufficient.''}

As illustrated in \cref{fig_loss}, forward learning converges rapidly but tends to overfit the backbone model, whereas backward learning mitigates this limitation. Motivated by this observation, our cascade learning proceeds by performing forward learning followed by backward learning. Importantly, we find that only one epoch is required for each stage to reach optimal performance. This design directly supports one of our main contributions: highly efficient computation.

As reported in \cref{tab_minimalsufficient}, a single gradient computation on a CIFAR-10 image takes approximately 0.01 seconds on an NVIDIA A100 GPU (with one learning epoch and without SmoothGrad). While alternative configurations, such as one-epoch learning alone or variants with or without gradient smoothing, exhibit inferior performance compared to our proposed approach (gray-highlighted), the cascade strategy achieves near-optimal results under this minimal setting. Increasing the number of training epochs yields only marginal gains, as evidenced by the 4-epoch cascade configuration.

\begin{table}[!t]
    \scriptsize
    \centering
    \begin{threeparttable}
        \caption{\revise{Number of epochs in cascade learning.}}
        \label{tab_minimalsufficient}
        \setlength{\tabcolsep}{3.4mm}{\begin{tabular}{ccccc}
            \toprule
            Learning&SmoothGrad&Time (s)&Clean&Robust\\
            \midrule
            Forward (1 epoch)&N/A&0.01&95.4&86.1\\
            Forward (1 epoch)&20&0.01&95.2&86.5\\
            Backward (1 epoch)&N/A&0.01&95.2&85.8\\
            Backward (1 epoch)&20&0.01&95.2&86.5\\
            Cascade (2 epoch)&N/A&0.02&95.7&87.4\\
            \rowcolor{gray!10}Cascade (2 epoch)&20&0.02&95.9&88.1\\
            Cascade (4 epoch)&20&0.04&95.9&88.2\\
            \bottomrule
        \end{tabular}}
         \begin{tablenotes}
            \item The gray-highlighted row denotes the proposed algorithm and its corresponding configuration. The attack is CIFAR-10, $\ell_\infty$-bounded PGD attacks with $\epsilon = 8/255$, and the target model is WideResNet-94-16 \cite{bartoldson2024adversarial}.
        \end{tablenotes}
    \end{threeparttable}
\end{table}

\textbf{Minimal.}
The proposed MSPD and MCGS are termed \emph{``minimal''} because the learning process consists of only two cascade epochs: one forward and one backward. This represents the minimal configuration required to constitute cascade learning; otherwise, the process would reduce to purely forward-only or backward-only learning. Moreover, within each epoch, MCGS replaces iterative gradient descent methods (e.g., PGD) with a single-step gradient smoothing operation, further reducing computational overhead and enabling efficient, parallelized gradient computation.

\textbf{Sufficient.}
Despite its simplicity, this minimal configuration is \emph{``sufficient''} to achieve competitive robustness. Empirical results demonstrate that the two-epoch cascade already reaches near-optimal performance, while additional epochs provide negligible improvements.

\end{reviseG}

\subsection{Single-step Gradient Smoothing vs. Iterative Gradient Descent}
Gradient smoothing improves efficiency and generalization over iterative descent. Results in \cref{tab_ablation_smoothgrad} show our defense achieves near-optimal performance with just two epochs, the minimal cost for cascade learning.

\begin{table}[!t]
    \scriptsize
    \centering
    \begin{threeparttable}
        \caption{Gradient Smoothing \vs PGD.}
        \label{tab_ablation_smoothgrad}
        \setlength{\tabcolsep}{2.3mm}{\begin{tabular}{ccccc}
            \toprule
            Algorithm&$N$&Time (s)&Clean (\%)&Robust (\%)\\
            \midrule
            \multirow{2}{*}{PGD20}&N/A&0.49&95.9&87.7\\
            &20&0.48&95.7&88.1\\
            \midrule
            \multirow{3}{*}{\begin{tabular}{c}Gradient Smoothing\\ (\textbf{Ours})\end{tabular}}&N/A&0.02&95.7&87.4\\
            &10&0.02&95.9&88.0\\
            &20 (\textbf{Ours})&0.02&95.9&88.1\\
            \bottomrule
        \end{tabular}}
        \begin{tablenotes}
            \item Single-step gradient smoothing resists gradient obfuscation from unseen models and attacks more effectively than increasing per-epoch iterations.
        \end{tablenotes}
    \end{threeparttable}
\end{table}

\subsection{Step Size}
\cref{tab_ablation_alpha} shows the effect of step size $\alpha$ in single-step gradient smoothing under a fixed two-epoch budget. All settings incur similar time, but higher $\alpha$ improves both clean and robust accuracy. Our default $\alpha = 1.00$ yields the best results, supporting that $\alpha \times \mathrm{Epochs} > 1$ ensures sufficient update magnitude.

\begin{table}[!t]
    \scriptsize
    \centering
    \begin{threeparttable}
        \caption{Step size $\alpha$.}
        \label{tab_ablation_alpha}
        \setlength{\tabcolsep}{5.7mm}{\begin{tabular}{cccc}
            \toprule
            Step Size $\alpha$&Time (s)&Clean (\%)&Robust (\%)\\
            \midrule
            0.50&0.02&95.5&86.8\\
            0.75&0.02&95.8&88.0\\
            1.00 (\textbf{Ours})&0.02&95.9&88.1\\
            \bottomrule
        \end{tabular}}
        \begin{tablenotes}
            \item Learning requires $\alpha \times \mathrm{Epochs} > 1$ to ensure sufficient update magnitude.
        \end{tablenotes}
    \end{threeparttable}
\end{table}

\begin{table}[!t]
    \scriptsize
    \centering
    \begin{threeparttable}
        \caption{Fixed defense budget \vs varying attack budgets.}
        \label{tab_ablation_eps}
        \setlength{\tabcolsep}{4.4mm}{\begin{tabular}{cccc}
            \toprule
            Attack $\epsilon$&Defense $\epsilon$&No Defense (\%)&With Defense (\%)\\
            \midrule
            16/255&8/255&38.0&60.3\\
            8/255&8/255&77.1&88.1\\
            4/255&8/255&87.9&92.8\\
            \bottomrule
        \end{tabular}}
        \begin{tablenotes}
            \item With fixed defense, our method resists stronger attacks and outperforms no defense.
        \end{tablenotes}
    \end{threeparttable}
\end{table}

\subsection{Robustness Under Stronger Attack Settings}
\label{ablation_eps}
In realistic deployments, defenses must remain fixed while facing adversarial attacks of varying and potentially unknown strengths. Unlike certified defenses, which typically provide guarantees only under small perturbation budgets (\eg $\epsilon \leq 8/255$ in $\ell_\infty$ norm), preemptive defenses must remain effective even when adversaries exceed these bounds. To assess this, we evaluate MSPD under both weaker and stronger perturbation magnitudes. As shown in \cref{tab_ablation_eps}, our defense consistently maintains high clean and robust accuracy across a wide range of $\epsilon$, including challenging settings such as $\epsilon = 16/255$. Notably, robustness margins often improve with increasing $\epsilon$, indicating that MSPD generalizes well under fixed configurations even in high-threat scenarios. These results suggest that while MSPD does not offer certified guarantees, it achieves practical robustness beyond the operational limits of certification-based methods.

\begin{reviseG}
\section{Discussion}

\begin{table}[!t]
    \scriptsize
    \centering
    \begin{threeparttable}
        \caption{\revise{Preemptive defenses for nonrobust models.}}
        \label{tab_nonrobust}
        \setlength{\tabcolsep}{1.1mm}{\begin{tabular}{cc|cc|cc|cc}
            \toprule
            \multirow{4}{*}{Defense}&\multirow{4}{*}{Backbone}&\multicolumn{6}{c}{Target}\\
            &&\multicolumn{2}{c|}{\makecell{XCiT-L12\\(Robust)}}&\multicolumn{2}{c|}{\makecell{WideResNet-28-10\\(Nonrobust)}}&\multicolumn{2}{c}{\makecell{ResNet-56\\(Nonrobust)}}\\
            &&Clean&Robust&Clean&Robust&Clean&Robust\\
            \midrule
            No Pre.&N/A&92.8&60.2&94.8&0.0&94.3&2.5\\
            \midrule
            UnAdv&\makecell{UnAdv (Robust)}&91.8&61.7&93.2&0.0&91.4&2.1\\
            Bi-Level&\makecell{Bi-Level (Robust)}&91.5&72.2&93.8&0.0&92.8&2.8\\
            \textbf{Ours}&\makecell{\textbf{Ours} (Robust)}&\textbf{95.3}&\textbf{78.7}&\textbf{95.7}&\textbf{0.4}&\textbf{95.3}&\textbf{4.0}\\
            \midrule
            \textbf{Ours}&\makecell{WideResNet-28-10\\(Nonrobust)}&92.6&61.2&\cellcolor{gray!10}95.8&\cellcolor{gray!10}0.6&95.1&2.7\\
            \textbf{Ours}&\makecell{ResNet-56\\(Nonrobust)}&92.8&61.0&94.9&0.0&\cellcolor{gray!10}95.2&\cellcolor{gray!10}4.3\\
            \bottomrule
        \end{tabular}}
        \begin{tablenotes}
            \item \revise{Gray denotes the white-box setting (identical backbone and target). \textbf{Bold} indicates the best transferable performance. Attacks use CIFAR-10 $\ell_\infty$-bounded PGD with $\epsilon = 8/255$. Although robustness gains on nonrobust targets are limited, our method consistently improves clean accuracy across unseen models, including nonrobust ones.}
        \end{tablenotes}
    \end{threeparttable}
\end{table}

\begin{table}[!t]
\scriptsize
\centering
\begin{threeparttable}
\caption{\revise{Performance under different defense combinations.}}
\label{tab_combination}
\setlength{\tabcolsep}{5.9mm}
\begin{tabular}{r|cc}
\toprule
Combination&Clean&Robust\\
\midrule
Training-time&93.9&77.1\\
Preemptive + Training-time&\textbf{95.9}&\textbf{88.1}\\
\midrule
Test-time&89.2&87.5\\
Preemptive + Test-time&\textbf{94.4}&\textbf{93.3}\\
\midrule
Test-time + Training-time&91.9&76.5\\
Preemptive + Test-time + Training-time&\textbf{95.5}&\textbf{89.2}\\
\bottomrule
\end{tabular}
\end{threeparttable}
\begin{tablenotes}
    \item \revise{Training-time defense: WideResNet-94-16~\cite{bartoldson2024adversarial}; test-time defense: DiffPure~\cite{nie2022diffusion}. \textbf{Bold} indicates the best performance. Adding our preemptive defense consistently improves layered defense combinations in both clean and robust accuracy.}
\end{tablenotes}
\end{table}

\subsection{Protecting Nonrobust Models}
\cref{tab_nonrobust} identifies a boundary condition of preemptive defenses, including MSPD: they do not substantially improve adversarial robustness on nonrobust target models, even under white-box alignment (identical backbone and target). In contrast, large gains are observed when defending robust targets. This limitation stems from the nature of nonrobust models rather than surrogate–target mismatch. Their adversarial solution space is highly unconstrained, allowing many valid perturbations within the same budget. In such a regime, a single preemptively learned perturbation pattern can cover only a limited subset of adversarial directions, restricting achievable robustness even under perfect backbone alignment.

Importantly, MSPD consistently improves \emph{clean accuracy} across unseen target models, including nonrobust ones, while existing preemptive baselines do not. Moreover, when combined with other defenses (see \Cref{tab_combination}), MSPD significantly enhances both clean and robust accuracy. For example, adding MSPD to a robust training-time model improves robust accuracy from 77.1\% to 88.1\%, and incorporating it into a layered test-time pipeline on a nonrobust model (WideResNet-28-10) increases robustness from 87.5\% to 93.3\%.

These results clarify MSPD’s role: it is not a standalone solution for unconstrained nonrobust models, but a lightweight and effective reinforcement mechanism. When layered with training-time or test-time defenses, it consistently strengthens overall system robustness at negligible computational cost. Extending preemptive defenses to fully unconstrained adversarial regimes remains an important direction for future work.

\subsection{Resisting White-Box Adaptive Reversion}

\cref{performance_reversion} shows that effective reversion occurs only under a fully white-box stress test with exact backbone access. Even in this strongest setting, MSPD still improves clean accuracy by $+0.5\%$ and robust accuracy by $+2.2\%$ over no defense.

\textbf{Backbone availability.} Public robust models (e.g., RobustBench) are convenient but accessible. However, an attacker could also exploit them to facilitate white-box reversion attacks. Moreover, for certain tasks, suitable public robust models may not be available. A more secure alternative is for the defender to train a private backbone. Although adversarial training is computationally expensive, our results in \cref{ablation_backbone} show that a lightweight robust backbone is sufficient. This substantially reduces the training cost, though private adversarial training remains a practical limitation.

\textbf{Realism of white-box reversion.}
Using a public backbone does not automatically imply the idealized white-box setting studied here. Successful reversion requires reproducing the defender’s exact checkpoint, preprocessing, and hyperparameters. In practice, even if an attacker can enumerate candidate public backbones, identifying the defender’s exact instantiation is non-trivial without privileged information. Because perturbations are near-imperceptible (\cref{qualitative}), visual inspection is insufficient; attackers would need repeated system interaction under practical constraints (e.g., query limits or stochastic inference). These factors make exhaustive reversion substantially harder than the stress-test scenario, consistent with the failure of gray-/black-adaptive attacks (\cref{fig_reversion}).

\subsection{On the Practicality of Adaptive Reversion Attacks}
\label{sec:adaptive_reversion_limits}

Preemptive Reversion is a diagnostic stress test probing the upper bound of attacker capability. Its success requires precise gradient access and strong model alignment. Non-gradient adaptive strategies, including JPEG compression and DiffPure~\cite{nie2022diffusion}, fail to reverse our defense (\cref{lemma_reversion}, \cref{fig_reversion}) because our perturbations are semantically aligned rather than noise-like. Without closely aligned gradients, perturbation recovery is ill-posed. In gray- and black-adaptive settings, misalignment often amplifies rather than cancels the defense. Effective reversion is therefore observed only under the fully white-box assumption with exact backbone access—a pessimistic upper bound on attacker power. Outside this regime, adaptive reversion is ineffective, reinforcing its interpretation as a worst-case diagnostic rather than a practical threat model.

\end{reviseG}

\section{Conclusion}
We proposed a novel preemptive defense, MSPD, that neutralizes adversarial threats before attack time. At its core is the MCGS algorithm, which achieves strong clean and robust accuracy with minimal cost. We analyze the roles of the backbone and classifier, and introduce a white-box adaptive diagnostic showing that successful reversal requires impractical gradient access. Extensive experiments across datasets, threat models, and unseen architectures confirm SOTA robustness, visual quality, and speed. Ablations further validate key choices, including the classifier, backbone, and algorithms. Future work will extend MSPD to multi-label and multimodal inputs. Moreover, defenses remain vulnerable under full transparency to adaptive attackers, making it particularly challenging to design adaptive-aware strategies that can genuinely withstand such threats.

{
    \bibliographystyle{ieeetr}
    \bibliography{main}
}
\begin{IEEEbiography}[{\includegraphics[width=1in,height=1.25in,clip,keepaspectratio]{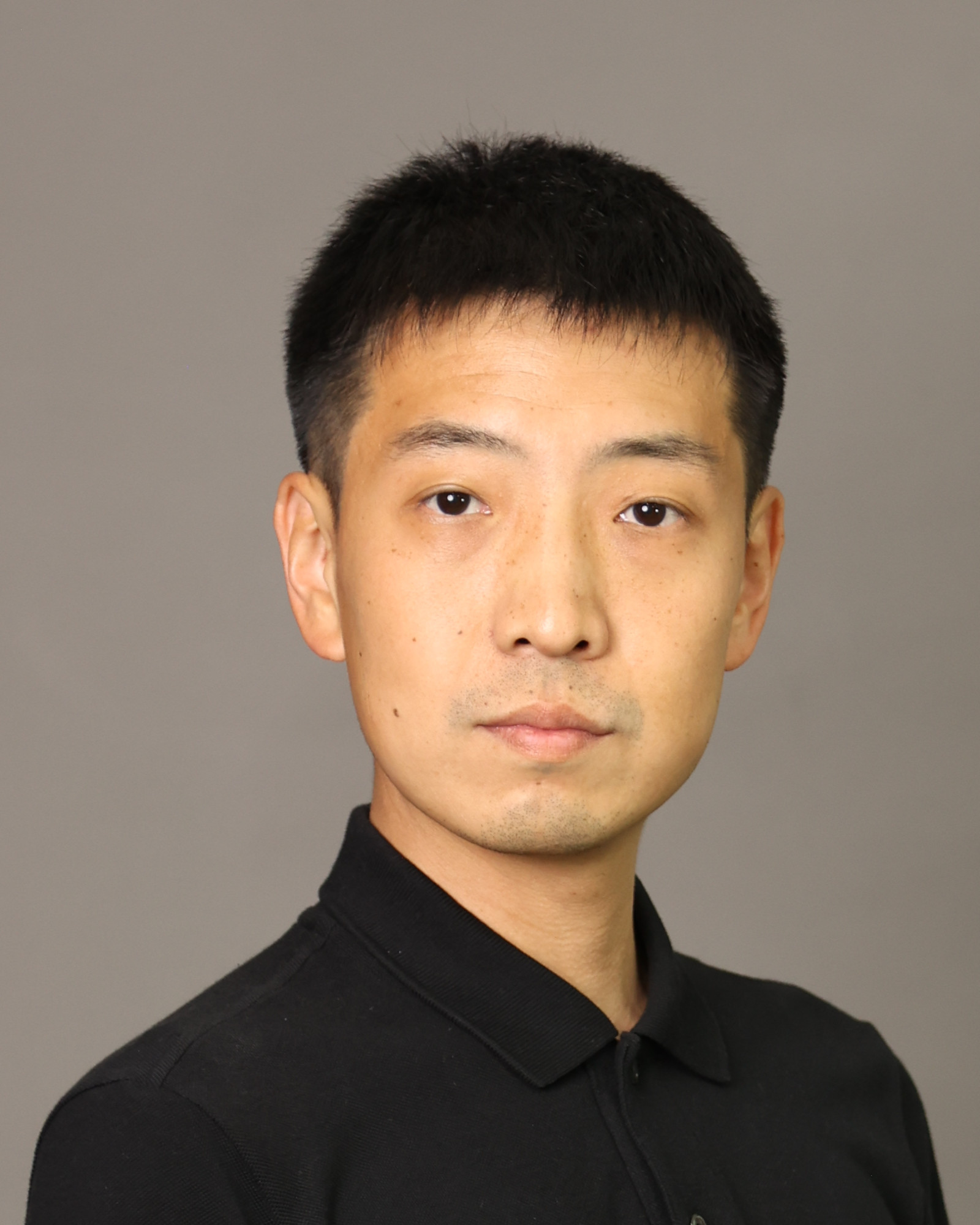}}]{Hanrui Wang}
received his B.S. degree in Electronic Information Engineering from Northeastern University (China) in 2011. After working in the IT industry and rising to a director-level position, he transitioned to academic research in 2019 and earned his Ph.D. in Computer Science from Monash University, Australia, in January 2024. He is currently an Assistant Professor in the Echizen Laboratory at the National Institute of Informatics (NII) in Tokyo, Japan. His research focuses on AI security and privacy, with emphasis on adversarial machine learning, model inversion, and trustworthy multimedia systems. He has published multiple papers in leading journals such as IEEE TIFS, IEEE TDSC, ACM TOMM, and J-STARS, as well as in reputable international conferences including WWW, WACV, FG, ICPR, ICASSP, and ICMI.
\end{IEEEbiography}

\begin{IEEEbiography}[{\includegraphics[width=1in,height=1.25in,clip,keepaspectratio]{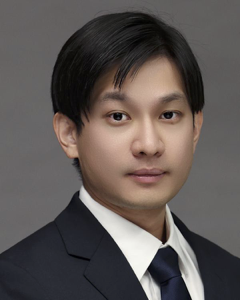}}]{Ching-Chun Chang} (Senior Member, IEEE) received the PhD degree in Computer Science from the University of Warwick, UK, in 2019. He is currently affiliated with the National Institute of Informatics, Japan, as a Project Assistant Professor. He also serves as a Visiting Researcher at Peking University, China, and a Distinguished Professor at Hangzhou Dianzi University, China. He participated in a Short-Term Scientific Mission supported by European Cooperation in Science and Technology Actions at the Faculty of Computer Science, Otto von Guericke University of Magdeburg, Germany, in 2016. He was granted the Marie-Curie Fellowship and participated in a Research and Innovation Staff Exchange scheme supported by Marie Skłodowska-Curie Actions at the Department of Electrical and Computer Engineering, New Jersey Institute of Technology, USA, in 2017. He was a Visiting Scholar with the School of Computing and Mathematics, Charles Sturt University, Australia, in 2018, and with the School of Information Technology, Deakin University, Australia, in 2019. He was a Research Fellow with the Department of Electronic Engineering, Tsinghua University, China, in 2020. His research interests include artificial intelligence, biometrics, cryptography, cybersecurity, evolutionary computation, forensics, information theory, steganography, and watermarking.
\end{IEEEbiography}

\begin{IEEEbiography}[{\includegraphics[width=1in,height=1.25in,clip,keepaspectratio]{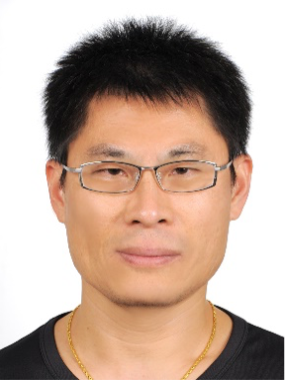}}]{Chun-Shien Lu}
received the Ph.D. degree in Electrical Engineering from National Cheng-Kung University, Tainan, Taiwan in 1998. He is a full research fellow (full professor) in the Institute of Information Science since March 2013 and the executive director in the Taiwan Information Security Center, Research Center for Information Technology Innovation, Academia Sinica, Taipei, Taiwan, since June 2024. His current research interests mainly focus on deep learning, AI security and privacy, and inverse problems. Dr. Lu serves as a Technical Committee member of Communications and Information Systems Security (CIS-TC) and Multimedia Communications Technical Committee (MMTC), IEEE Communications Society, since 2012 and 2017, respectively. Dr. Lu also serves as Area Chairs of ICASSP 2012--2014, ICIP 2013, ICME 2018, ICIP 2019--2025, ICML 2020, ICML 2023--2025, ICLR 2021--2026, NeurIPS 2022--2025, and ACM Multimedia 2022--2025, and Senior program committee of AAAI 2025--2026. Dr. Lu has owned four US patents, five ROC patents, and one Canadian patent in digital watermarking and graphic QR code. Dr. Lu won Ta-You Wu Memorial Award, National Science Council in 2007 and was a co-recipient of a National Invention and Creation Award in 2004. Dr. Lu was an associate editor of IEEE Trans. on Image Processing from 2010/12 to 2014 and from 2018/3 to 2023/6.
\end{IEEEbiography}

\begin{IEEEbiography}[{\includegraphics[width=1in,height=1.25in,clip,keepaspectratio]{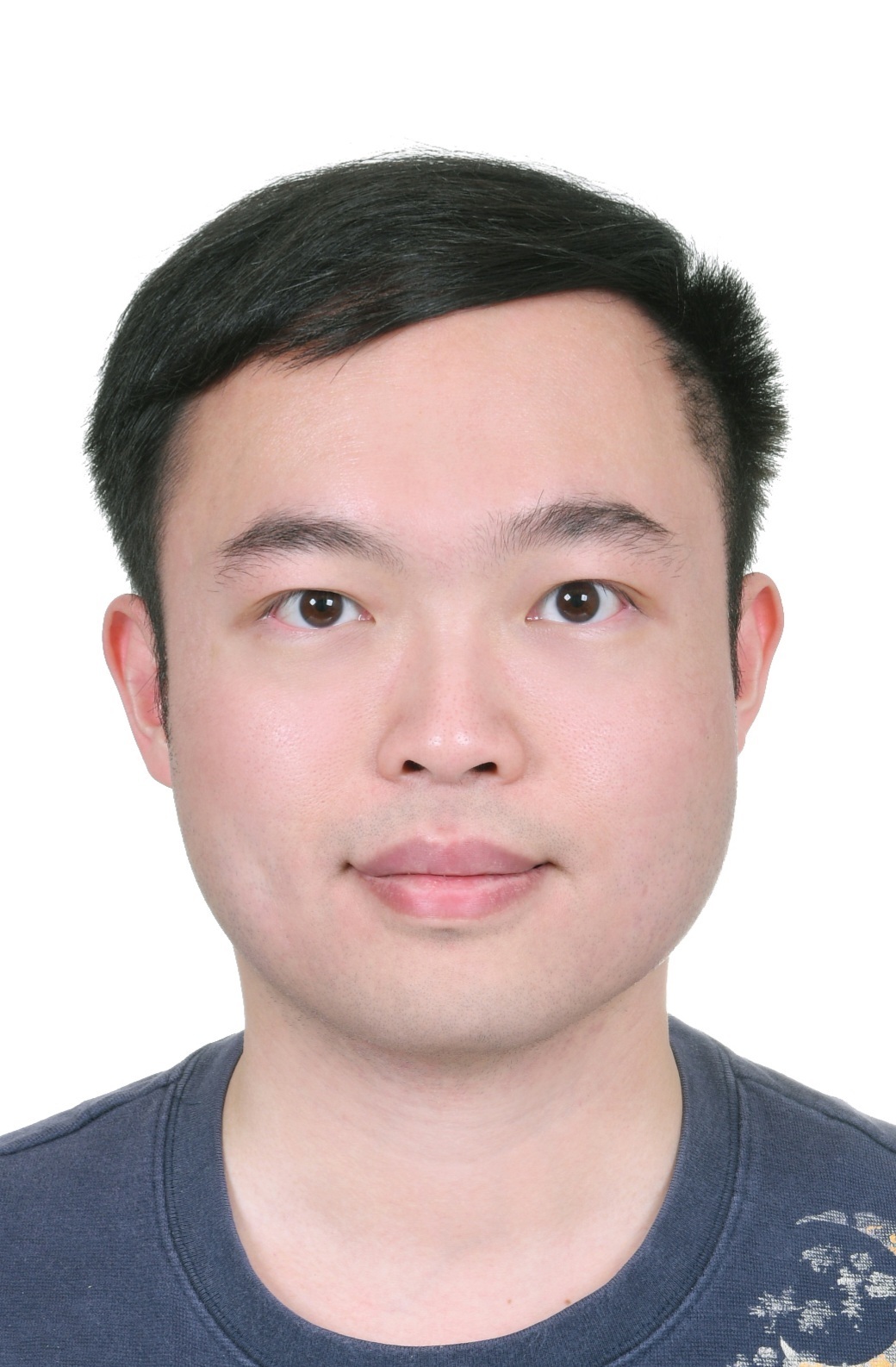}}]{Ching-Chia Kao} received the M.S. degree in Applied Mathematics from National Yang Ming Chiao Tung University, Hsinchu, Taiwan, in 2018. He is currently a Ph.D. student in the Department of Computer Science and Information Engineering, National Taiwan University, Taipei, Taiwan. He is also a Research Assistant with the Institute of Information Science and the Research Center for Information Technology Innovation, Academia Sinica, Taipei, Taiwan. His research interests include adversarial machine learning, differential privacy, and LLM jailbreak attacks.
\end{IEEEbiography}

\begin{IEEEbiography}[{\includegraphics[width=1in,height=1.25in,clip,keepaspectratio]{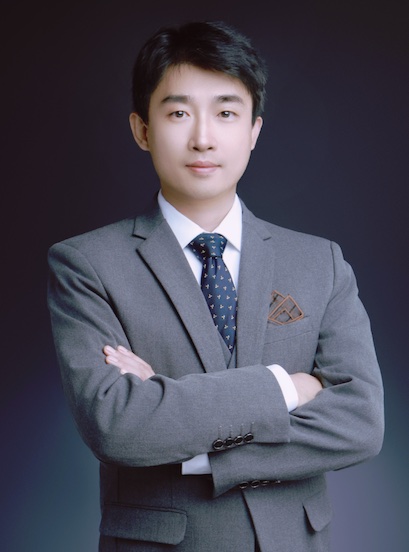}}]{Shuo Wang}  (Senior Member, IEEE) is an Associate Professor at Shanghai Jiao Tong University, specializing in cybersecurity and artificial intelligence. Before joining SJTU, he served as a Senior Research Scientist at the Commonwealth Scientific and Industrial Research Organisation (CSIRO), Australia’s national science agency. He received his Ph.D. in Computer Science from the University of Melbourne in June 2018. Dr. Wang serves as an Associate Editor for IEEE Transactions on Dependable and Secure Computing (TDSC) and IEEE Transactions on Information Forensics and Security (TIFS). His research focuses on Responsible and Trustworthy AI, particularly on enhancing the robustness, interpretability, and reliability of deep learning and graph-based models. He is also dedicated to addressing critical challenges in cybersecurity and privacy across systems, networks, and data-driven infrastructures. Dr. Wang has published more than 60 research papers in prestigious journals and top-tier conferences in information security and artificial intelligence, including IEEE S\&P, NDSS, USENIX Security, CCS, CVPR, ICML, ICLR, IEEE TIFS, IEEE TDSC, IEEE TPDS, IEEE TSC, IEEE TNNLS, WWW, and ESEC/FSE.
\end{IEEEbiography}

\begin{IEEEbiography}[{\includegraphics[width=1in,height=1.25in,clip,keepaspectratio]{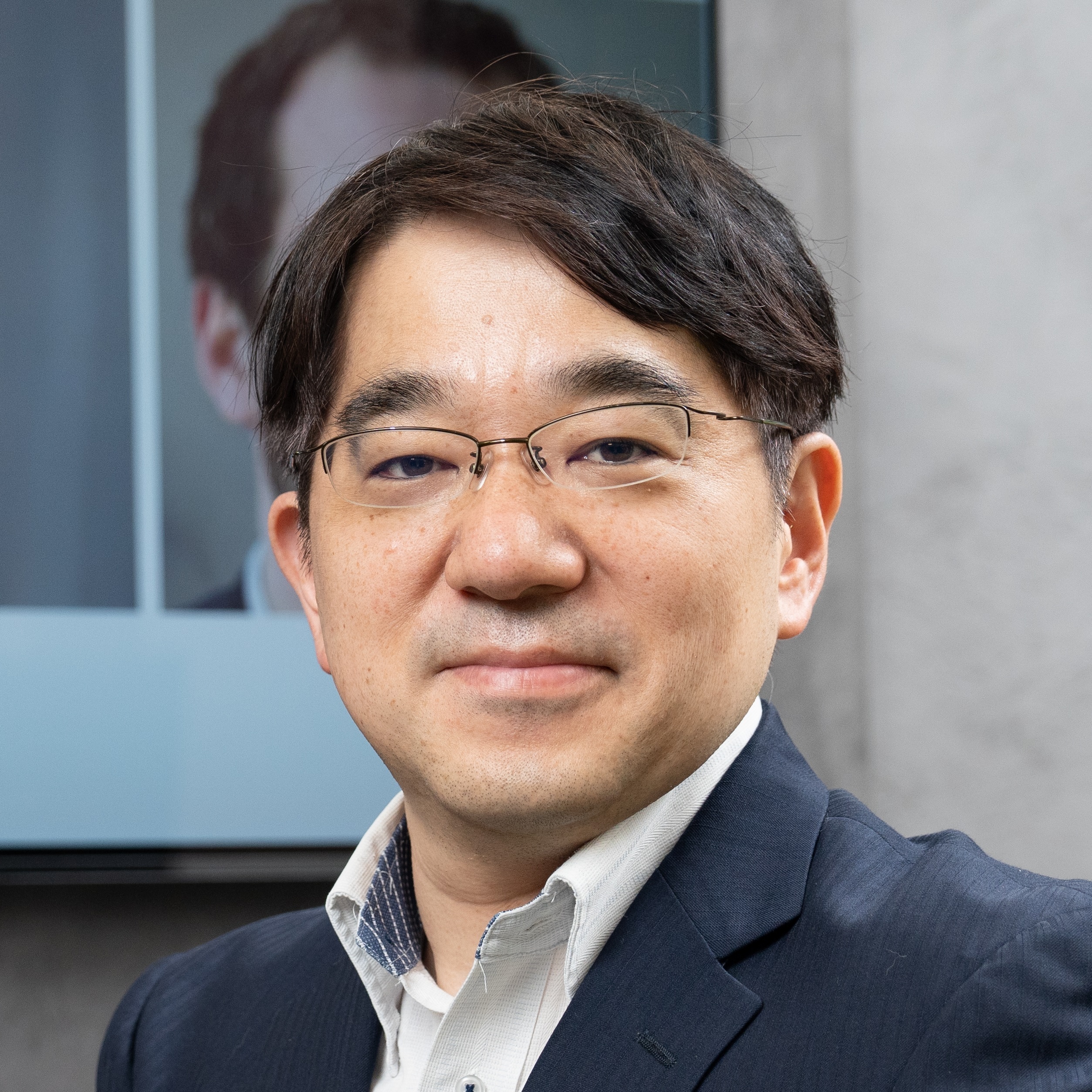}}]{Isao Echizen} (Senior Member, IEEE) received B.S., M.S., and D.E. degrees from the Tokyo Institute of Technology, Japan, in 1995, 1997, and 2003, respectively. He joined Hitachi, Ltd. in 1997 and until 2007 was a research engineer in the company's systems development laboratory. He is currently a director and a professor of the Information and Society Research Division, the National Institute of Informatics (NII), a director of the Global Research Center for Synthetic Media, the NII, a professor in the Department of Information and Communication Engineering, Graduate School of Information Science and Technology, the University of Tokyo, and a professor in the Graduate Institute for Advanced Studies, the Graduate University for Advanced Studies (SOKENDAI), Japan.  He was a visiting professor at the University of Freiburg, Germany, and at the University of Halle-Wittenberg, Germany. He is currently engaged in research on AI security, multimedia security, and multimedia forensics. He is a research director in the CREST FakeMedia project and in the K Program SYNTHETIQ X, Japan Science and Technology Agency (JST). He received the Commendation for Science and Technology by the Minister of Education, Culture, Sports, Science and Technology (Research category) in 2025, the Best Paper Award from the IEICE in 2023, the Best Paper Awards from the IPSJ in 2005 and 2014, the IPSJ Nagao Special Researcher Award in 2011, the DOCOMO Mobile Science Award in 2014, the Information Security Cultural Award in 2016, and the IEEE Workshop on Information Forensics and Security Best Paper Award in 2017. He was a member of the Information Forensics and Security Technical Committee of the IEEE Signal Processing Society. He is the IEICE Fellow, the IPSJ Fellow, the IEEE Senior Member, and the Japanese representative on IFIP and on IFIP TC11 (Security and Privacy Protection in Information Processing Systems), a vice president of APSIPA, and an editorial board member of the IEEE Transactions on Dependable and Secure Computing, the EURASIP Journal on Image and Video Processing, and the Journal of Information Security and Applications, Elsevier.
\end{IEEEbiography}
\clearpage
\appendices

\section{Proof}
\label{proof}
\setcounter{lemma}{0}
\begin{lemma}
\label{lemma_backbone_appx}
Let $f_{\mathrm{back}}$ and $f_{\mathrm{tgt}}$ denote backbone and target models. If $\nabla_x f_{\mathrm{back}}(x)$ and $\nabla_x f_{\mathrm{tgt}}(x)$ share a vulnerable subspace, then defense perturbations $\delta^{\mathrm{def}}$ computed from $f_{\mathrm{back}}$ transfer to $f_{\mathrm{tgt}}$ and suppress future attacks.
\end{lemma}

This condition favors robust backbones, which are known to yield semantically meaningful and architecture-agnostic gradients~\cite{tramer2018ensemble}. We adopt adversarially trained models from model zoos~\cite{croce2021robustbench}, as they empirically satisfy the lemma’s hypothesis and improve transfer across diverse architectures. The empirical evidence with success/failure case analysis are provided in \Cref{tab_ablation_backbone} and \Cref{tab_ablation_pipeline}.

\begin{proof}
Let $x \in \mathcal{X}$ be an input with ground-truth label $y$. Let $g_{\mathrm{back}}(x),\ g_{\mathrm{tgt}}(x) \in \mathbb{R}^d$ be the input gradients of the backbone and target models, respectively, defined as
\begin{equation}
  \begin{aligned}
    g_{\mathrm{back}}(x) &= \nabla_x \mathcal{L}(f_{\mathrm{back}}(x), y),\\
    g_{\mathrm{tgt}}(x) &= \nabla_x \mathcal{L}(f_{\mathrm{tgt}}(x), y),
  \end{aligned}
\end{equation}
and let $\theta(x)$ denote the angle between them, with cosine similarity given by
\begin{equation}
    \cos\theta(x) = \frac{g_{\mathrm{back}}(x)^\top g_{\mathrm{tgt}}(x)}{\|g_{\mathrm{back}}(x)\|_2 \cdot \|g_{\mathrm{tgt}}(x)\|_2}.
\end{equation}
We assume throughout that both gradients are non-zero. Note that the input gradient of a classifier reflects how its prediction score changes with respect to input pixels; thus, directions in input space aligned with the gradient correspond to directions the model is most sensitive to.

\textbf{a. Shared Vulnerable Subspace:}
We prove the existence of a shared subspace when $\cos\theta(x) > 0$, such that any perturbation with non-zero projection into this space will influence both the backbone and target models.

Define the projection of $g_{\mathrm{back}}(x)$ onto $g_{\mathrm{tgt}}(x)$ as
\begin{equation}
    v(x) = \Pi_{ g_{\mathrm{tgt}}(x)} g_{\mathrm{back}}(x) 
    = \left( \frac{g_{\mathrm{tgt}}(x)^\top g_{\mathrm{back}}(x)}{\|g_{\mathrm{tgt}}(x)\|_2^2} \right) g_{\mathrm{tgt}}(x).
\end{equation}
Since both gradients are non-zero and $\cos\theta(x) > 0$, we have
\begin{equation}
    g_{\mathrm{tgt}}(x)^\top g_{\mathrm{back}}(x) > 0,
\end{equation}
which implies \( v(x) \neq 0 \), indicating that $g_{\mathrm{back}}(x)$ and $g_{\mathrm{tgt}}(x)$ are positively aligned and share a non-zero directional component. We therefore define the \emph{shared vulnerable subspace} as
\begin{equation}
    \mathcal{S}(x) = \operatorname{span}\{v(x)\} \subseteq \mathbb{R}^d,
    \label{eq_sharespace}
\end{equation}
which is a nontrivial one-dimensional subspace shared by both models. Here, $\operatorname{span}\{v(x)\}$ denotes the set of all scalar multiples of $v(x)$:
\begin{equation}
    \operatorname{span}\{v(x)\} = \left\{ \alpha \cdot v(x) \mid \alpha \in \mathbb{R} \right\}.
\end{equation}
Intuitively, $\mathcal{S}(x)$ captures a direction in input space to which both models are sensitive, even if their gradients differ overall.

Assume existing perturbation \( \delta \in \mathbb{R}^d \) and its projection onto $S(x)$ satisfies \( \Pi_{\mathcal{S}} \delta \neq 0 \). Since \( \mathcal{S}(x) \subseteq \operatorname{span}\{g_{\mathrm{back}}(x)\} \) and \( \mathcal{S}(x) \subseteq \operatorname{span}\{g_{\mathrm{tgt}}(x)\} \), it follows that
\begin{equation}
    g_{\mathrm{back}}(x)^\top \Pi_{\mathcal{S}} \delta \neq 0,\quad g_{\mathrm{tgt}}(x)^\top \Pi_{\mathcal{S}} \delta \neq 0,
\end{equation}
so this component $\delta \in \mathbb{R}^d \rightarrow \Pi_{\mathcal{S}} \delta \neq 0$ affects the output of both $f_{\mathrm{back}}$ and $f_{\mathrm{tgt}}$. Hence, a shared direction exists in which perturbations influence both models, enabling transferability across architectures.

\textbf{b. Transferable Defense:}
Let $\delta_{\mathrm{back}} \in \mathbb{R}^d$ be an $\ell_p$-bounded perturbation generated to maximize the loss of $f_{\mathrm{back}}$, \ie an adversarial perturbation from the backbone model. Although $\delta_{\mathrm{back}}$ is adversarial for $f_{\mathrm{back}}$, it may contain transferable components that affect $f_{\mathrm{tgt}}$ via the shared subspace $\mathcal{S}(x)$.

We decompose $\delta_{\mathrm{back}}$ as
\begin{equation}
    \delta_{\mathrm{back}} = 
    \Pi_{\parallel} \delta_{\mathrm{back}} + 
    \Pi_{\perp} \delta_{\mathrm{back}},
\end{equation}
where $\Pi_{\parallel} \delta_{\mathrm{back}}$ is the projection onto $\mathcal{S}(x)$, and $\Pi_{\perp} \delta_{\mathrm{back}}$ is the orthogonal residual.

To construct a defense, we reverse the shared-space component to cancel future adversarial effects in that direction. For any scalar $\tau \in (0,1]$, we define the defense perturbation as:
\begin{equation}
    \delta^{\mathrm{def}} = -\tau \Pi_{\parallel} \delta_{\mathrm{back}},
\end{equation}
where $\tau$ is a tunable factor ensuring $\|\delta^{\mathrm{def}}\|_p \le \epsilon$, the allowed perturbation budget.

Because the shared subspace $\mathcal{S}(x)$ lies in $\operatorname{span}\{g_{\mathrm{tgt}}(x)\}$, we define the \emph{defense strength} as the inner product between $g_{\mathrm{tgt}}(x)$ and $\delta^{\mathrm{def}}$. Specifically, let $\theta(x)$ denote the angle between $g_{\mathrm{back}}(x)$ and $g_{\mathrm{tgt}}(x)$ as defined earlier. Then, since $\Pi_{\parallel} \delta_{\mathrm{back}}$ is aligned with $g_{\mathrm{tgt}}(x)$, we obtain:
\begin{equation}
    g_{\mathrm{tgt}}(x)^\top \delta^{\mathrm{def}}
    = -\tau \|g_{\mathrm{tgt}}(x)\|_2 \cdot 
    \|\Pi_{\parallel} \delta_{\mathrm{back}}\|_2 \cdot 
    \cos\theta(x) < 0.
\end{equation}
This expression shows that $\delta^{\mathrm{def}}$ moves $x$ in the direction opposite to the gradient of $f_{\mathrm{tgt}}$, reducing sensitivity to perturbations aligned with the shared space. We interpret $\delta^{\mathrm{def}}$ as a \emph{transferable defense}.

In practice, we do not isolate $\Pi_{\parallel} \delta_{\mathrm{back}}$ directly. Instead, we reverse the full perturbation $\delta_{\mathrm{back}}$, which implicitly includes the shared component. The defense is therefore most effective when a large portion of $\delta_{\mathrm{back}}$ lies within $\mathcal{S}(x)$.

Importantly, the effectiveness of the defense depends on the cosine similarity $\cos\theta(x)$ between the gradients. Suppose $\delta_{\mathrm{back}}$ is computed via gradient ascent on $f_{\mathrm{back}}$ (\eg FGSM and PGD), so that it aligns with $g_{\mathrm{back}}(x)$. Then, the projection onto the shared space satisfies
\begin{equation}
    \|\Pi_{\parallel} \delta_{\mathrm{back}}\|_2 
    = \|\delta_{\mathrm{back}}\|_2 \cdot \cos\theta(x).
\end{equation}
Substituting into the defense strength,
\begin{equation}
    g_{\mathrm{tgt}}(x)^\top \delta^{\mathrm{def}} 
    = -\tau \|\delta_{\mathrm{back}}\|_2 \cdot \|g_{\mathrm{tgt}}(x)\|_2 \cdot \cos^2\theta(x),
    \label{eq_target_negative}
\end{equation}
which shows that defense effectiveness scales quadratically with $\cos\theta(x)$.

Therefore, given fixed $\|\delta_{\mathrm{back}}\|_2$, a higher alignment between $g_{\mathrm{back}}(x)$ and $g_{\mathrm{tgt}}(x)$ directly yields a stronger protective effect.

\textbf{c. Success and Failure Cases:}
We now analyze why stronger backbone–target alignment increases the likelihood of successful defense on the target model. Let \( x^{\mathrm{def}} = x + \delta^{\mathrm{def}} \) be the defended input. Since all defenses begin from the same clean input \(x\), any difference in target-model robustness arises from the perturbation \( \delta^{\mathrm{def}} \), which depends on the alignment between \( f_{\mathrm{back}} \) and \( f_{\mathrm{tgt}} \).

\revise{The loss \( \mathcal{L}_{\mathrm{tgt}}(x) = \mathcal{L}(f_{\mathrm{tgt}}(x^{\mathrm{def}}), y) \) serves as a proxy for robustness:} lower loss suggests higher confidence and better resistance to attacks, though it does not guarantee success. As shown in \Cref{eq_target_negative} earlier, the defense perturbation reduces the loss on the target model when the backbone and target gradients are aligned.

Assuming the loss is \(L\)-smooth (following the standard assumption in Theorem~A.1 of~\cite{madry2018towards}), the first-order upper bound on target loss becomes
\begin{equation}
    \mathcal{L}_{\mathrm{tgt}}(x^{\mathrm{def}})
    \le
    \mathcal{L}_{\mathrm{tgt}}(x) + g_{\mathrm{tgt}}(x)^\top \delta^{\mathrm{def}} + \frac{L}{2} \|\delta^{\mathrm{def}}\|_2^2.
\end{equation}
Substituting the earlier result \( g_{\mathrm{tgt}}(x)^\top \delta^{\mathrm{def}} \) in \Cref{eq_target_negative}, the right-hand side decreases as \( \cos^2\theta(x) \) increases. Therefore, stronger alignment yields greater loss suppression on the target model.

While this loss decrease does not guarantee robustness, it shifts defended inputs into lower-loss regions that are typically more robust to attacks. Therefore, greater alignment implies:
\begin{itemize}
    \item a larger shared subspace \( \mathcal{S}(x) \),
    \item stronger descent transferred from backbone to target,
    \item and a higher probability that \( x^{\mathrm{def}} \) lies in a low-loss, robust region.
\end{itemize}

Conversely, if \(\cos\theta(x) \approx 0\), the reversed perturbation fails to reduce target loss meaningfully, and the defense becomes ineffective. Thus, backbone–target gradient alignment is a key factor governing whether a defended input becomes robust or vulnerable on the target model.

\textbf{d. Robust Models:}
\revise{Attacking adversarially trained models leads to more semantically consistent perturbations across different architectures~\cite{wang2025greedy} (see \Cref{fig_gradient_similarity}).} Let $\cos(\cdot,\cdot)$ denote cosine similarity between vectors. For robust models, there exists a constant $\rho_{\min} > 0$ such that
\begin{equation}
    \mathbb{E}_{(x,y)\sim\mathcal{D}}
    \left[
        \cos\big(g_{\mathrm{back}}(x),\ g_{\mathrm{tgt}}(x)\big)
    \right] \ge \rho_{\min},
    \label{eq:cosbound}
\end{equation}
which guarantees that, on average, the gradients of $f_{\mathrm{back}}$ and $f_{\mathrm{tgt}}$ are positively aligned across the same input. This non-trivial alignment implies the existence of a shared vulnerable subspace $\mathcal{S}(x)$, enabling transferable defense perturbations through reversed projections. Moreover, the magnitude of $\cos\theta(x)$ governs the cancellation strength, linking the robustness of adversarially trained models to the effectiveness of our defense.

\textbf{e. NonRobust Models:}
\revise{By contrast, standard (nonrobust) models often rely on brittle, non-semantic features~\cite{wang2025greedy} and therefore typically exhibit negligible gradient alignment with other models (see \Cref{fig_gradient_similarity}). As a result, they are ineffective as backbones:}
\begin{equation}
    \mathbb{E}_{(x,y)\sim\mathcal{D}}
    \left[
        \cos\big(g_{\mathrm{back}}(x),\ g_{\mathrm{tgt}}(x)\big)
    \right] \approx 0.
\end{equation}
This implies that the gradient fields are largely orthogonal and fail to induce any meaningful shared subspace. As a result, perturbations from nonrobust backbones have no transferable component and offer no protective effect when reversed. Empirically, we observe that such backbones result in $0\%$ robust accuracy under strong attacks, confirming their ineffectiveness.
\end{proof}

\begin{tcolorbox}[takeaway,title=Takeaway]
When the input-gradient fields of a robust backbone model and a target model are consistently aligned, a shared subspace $\mathcal{S}(x)$ emerges, enabling the backbone to compute perturbations that transfer to the target. By reversing the shared component of these perturbations, the backbone generates a defense that cancels out potential adversarial directions in the target model. The effectiveness of this defense is governed by the alignment strength, quantified by $\cos\theta(x)$, completing the proof of \Cref{lemma_backbone}.
\end{tcolorbox}


\begin{lemma}
\label{lemma_cascade_speed_appx}
Forward learning reduces loss more effectively than backward learning:
\begin{equation}
    \mathcal{L}(x) - \mathcal{L}(x_K^{\mathrm{F}}) 
    > 
    \mathcal{L}(x) - \mathcal{L}(x_K^{\mathrm{B}}).
\end{equation}
\end{lemma}

\noindent
This holds because forward learning follows direct gradient descent, while backward learning reverses accumulated ascent, which deviates from the optimal descent path.

\begin{proof}
Let \(K \ge 1\) denote the number of inner steps in a learning epoch. At each step \(k = 0, \dots, K-1\), we compute the input gradient of the backbone model:
\begin{equation}
g_k = \nabla_x \mathcal{L}(f_{\mathrm{back}}(x_k), \hat{y}),
\end{equation}
which we decompose into norm and direction:
\begin{equation}
g_k = s_k u_k, \quad \mathrm{where} \quad s_k = \|g_k\|_2, \quad u_k \in \mathbb{S}^{d-1}.
\label{eq_decompose}
\end{equation}
Here, \(s_k\) is the gradient magnitude and \(u_k\) is the unit direction in input space at step \(k\). We further define average gradient norm of the epoch:
\begin{equation}
\bar s = \frac{1}{K} \sum_{k=0}^{K-1} s_k. 
\label{eq_geometry}
\end{equation}

Let \(\mathcal{L}(x) = \mathcal{L}(f_{\mathrm{back}}(x), \hat{y})\) denote the loss function of the backbone model. We compare the total loss decrease over \(K\) steps of forward vs. backward learning, starting from the same input \(x_0 = x\).

\textbf{a. Forward learning:} performs gradient descent with step size \(\alpha > 0\), as defined in \Cref{eq_forward}. To quantify loss reduction in forward learning, we assume the loss function \( \mathcal{L} \) is differentiable and \(L\)-smooth, following the standard assumption in Theorem~A.1 of~\cite{madry2018towards};
that is, its gradient is Lipschitz continuous:
\begin{equation}
    \|\nabla \mathcal{L}(x) - \nabla \mathcal{L}(x')\| \le L \|x - x'\| \quad \forall x, x' \in \mathbb{R}^d.
    \label{eq_lsmooth}
\end{equation}
Under this assumption, the decrease in loss after a single forward gradient descent step is lower-bounded by:
\begin{equation}
    \mathcal{L}(x_k^{\mathrm{F}}) - \mathcal{L}(x_{k+1}^{\mathrm{F}}) 
    \ge 
    \alpha \left(1 - \frac{L\alpha}{2}\right) \|g_k^{\mathrm{F}}\|^2,
    \label{eq:stepwise_loss}
\end{equation}
as long as the step size satisfies \( \alpha < \frac{2}{L} \). This guarantees that each gradient step yields a predictable loss reduction proportional to the squared gradient norm.

Summing over all \(K\) steps and substituting the decomposition \(g_k = s_k u_k\) from \Cref{eq_decompose}, we obtain:
\begin{equation}
    \mathcal{L}(x) - \mathcal{L}(x_K^{\mathrm{F}})
    \ge 
    \alpha\left(1 - \frac{L\alpha}{2}\right) \sum_{k=0}^{K-1} {s_k^{\mathrm{F}}}^2.
    \label{eq:loss_fwd_scalar}
\end{equation}
This expression gives a lower bound on the total loss reduction after one forward learning epoch, governed by the accumulated squared gradient norms.

Using \Cref{eq_geometry}, we obtain a simplified lower bound for \Cref{eq:loss_fwd_scalar}:
\begin{equation}
    \mathcal{L}(x) - \mathcal{L}(x_K^{\mathrm{F}})
    \ge 
    K \alpha\left(1 - \frac{L\alpha}{2}\right) \bar{s^{\mathrm{F}}}^2.
    \label{eq:loss_fwd_avg}
\end{equation}
This shows that the total loss reduction scales linearly with the number of steps \(K\), and quadratically with the average gradient strength \(\bar{s^{\mathrm{F}}}\).

\textbf{b. Backward learning:} first constructs an adversarial example via gradient ascent, then reverses the perturbations, as defined in \Cref{eq_backward}. Now compare the original loss \(\mathcal{L}(x)\) and the final loss \(\mathcal{L}(x_K^{\mathrm{B}})\). \revise{Since the adversarial trajectory enters high-loss regions where gradients are noisy, the reverse point \(x_K^{\mathrm{B}}\) lies outside the clean descent manifold. The clean descent manifold refers to the region of the input space (under the clean loss) where gradient descent is stable and consistently converges to the low-loss clean data basin through direct optimization on the target model. In contrast, high-loss adversarial regions contain unstable gradients, causing optimization to wander rather than return to the clean basin.} Even under smoothness,
\begin{equation}
    \mathcal{L}(x) - \mathcal{L}(x_K^{\mathrm{B}})
    \le 
    K \alpha\left(1 - \frac{L\alpha}{2}\right) \bar{s^{\mathrm{adv}}}^2 \bar{c}.
    \label{eq:loss_back_avg}
\end{equation}
where $\bar{c}$ is the average cosine alignment between \(g_k^{\mathrm{adv}}\) and the forward descent direction \(-g_k^{\mathrm{F}}\) (\ie clean descent):
\begin{equation}
    \bar{c} = \frac{1}{K} \sum_{k=0}^{K-1} \cos\theta_k.
\end{equation}
In adversarial regions, these \(\cos\theta_k\) values are always less than one, even near zero due to incoherent, scattered gradients.
\end{proof}

\begin{tcolorbox}[takeaway,title=Takeaway]
Forward learning ensures monotonic loss descent proportional to the average gradient norm \(\bar{s^{\mathrm{F}}}\), whereas backward updates yield smaller and potentially unstable loss reductions due to:
\begin{itemize}
    \item Reduced alignment: \(\cos\theta_k < 1\) implies \(\bar{c} < 1\).
    \item Weaker gradients: adversarial regions exhibit lower average norm \(\bar{s^{\mathrm{adv}}} < \bar{s^{\mathrm{F}}}\), especially in robust models where adversarial directions often lie in flat regions, while clean inputs follow steeper paths toward semantically correct classes (empirical studies in \cite{ge2023boosting} measures lower $\|\nabla \mathcal{L}\|$ in adversarial flats).
    \item Suboptimal direction: backward updates reverse ascent steps, rather than following true descent.
\end{itemize}
Thus, under equal step size and budget, forward learning reduces loss more effectively:
\begin{equation}
    \mathcal{L}(x) - \mathcal{L}(x_K^{\mathrm{F}}) 
    > 
    \mathcal{L}(x) - \mathcal{L}(x_K^{\mathrm{B}}).
\end{equation}
\end{tcolorbox}


\revise{
\begin{lemma}
\label{lemma_cascade_transfer_appx}
Assume that the backward gradient components exhibit non-negligible average alignment with the shared vulnerable subspace $\mathcal{S}(x)$, i.e., their projections onto $\mathcal{S}(x)$ do not vanish in expectation. Then backward learning produces updates that place more energy in the shared vulnerable subspace than forward learning. Formally,
\begin{equation}
    \bigl\| \Pi_{\mathcal{S}} \Delta^{\mathrm{B}} \bigr\|_2^2 
    >
    \bigl\| \Pi_{\mathcal{S}} \Delta^{\mathrm{F}} \bigr\|_2^2 ,
\end{equation}
where $\Pi_{\mathcal{S}}$ denotes the orthogonal projection onto the shared vulnerable subspace $\mathcal{S}(x)$, and $\Delta^{\mathrm{*}}$ denotes the final perturbation generated by the corresponding update rule ($*$ being forward or backward). The projected norm $\|\Pi_{\mathcal{S}}\Delta^{*}\|_2^2$ measures how much of the perturbation lies in $\mathcal{S}(x)$, \ie the transferable component from the backbone model $f_{\mathrm{back}}$ to the target model $f_{\mathrm{tgt}}$.
\end{lemma}}

\noindent
Forward learning enforces “being right” under \( f_{\mathrm{back}} \), resulting in sharp, class-specific updates that transfer only when \( f_{\mathrm{back}} \approx f_{\mathrm{tgt}} \). In contrast, backward learning enforces “not being wrong,” producing more diffuse updates that broadly avoid errors, leading to more transferable robustness.

\begin{proof}
Let \( \mathcal{S}(x) \subseteq \mathbb{R}^d \) denote the shared vulnerable subspace between the backbone model \( f_{\mathrm{back}} \) and the target model \( f_{\mathrm{tgt}} \), as defined in \Cref{lemma_backbone}. For each training step \( k \in \{0,\dots,K-1\} \), define the input gradients:
\begin{equation}
    g_k^{\mathrm{F}} = \nabla_x \mathcal{L}(f_{\mathrm{back}}(x_k^{\mathrm{F}}), \hat{y}), \quad
    g_k^{\mathrm{B}} = \nabla_x \mathcal{L}(f_{\mathrm{back}}(x_k^{\mathrm{adv}}), \hat{y}),
\end{equation}
where \( x_k^{\mathrm{F}} \) are points on the forward descent trajectory, and \( x_k^{\mathrm{adv}} \) lie on the adversarial ascent path used in backward learning.

\textbf{a. Semantic-to-Geometric Interpretation:}
Forward learning enforces ``being right'' by increasing only the logit of the correct class \( y \). This leads to update directions that are nearly collinear:
\begin{equation}
    \mathcal{U}^{\mathrm{F}} = \operatorname{span}\{g_k^{\mathrm{F}}\}_{k=0}^{K-1}, \quad \operatorname{rank}(\mathcal{U}^{\mathrm{F}}) \approx 1.
\end{equation}
Backward learning enforces ``not being wrong'' by suppressing multiple incorrect logits, producing a trajectory that spans more directions:
\begin{equation}
    \mathcal{U}^{\mathrm{B}} = \operatorname{span}\{g_k^{\mathrm{B}}\}_{k=0}^{K-1}, \quad \operatorname{rank}(\mathcal{U}^{\mathrm{B}}) \gg 1.
\end{equation}

\textbf{b. Projection of Forward vs. Backward Updates:}
Let the final perturbations after \(K\) steps be:
\begin{equation}
    \Delta^{\mathrm{F}} = x_K^{\mathrm{F}} - x, \quad
    \Delta^{\mathrm{B}} = x_K^{\mathrm{B}} - x.
\end{equation}
We compare their projections onto the shared subspace \( \mathcal{S}(x) \). Let the backward update be decomposed into \( r \) orthogonal components:
\begin{equation}
    \Delta^{\mathrm{B}} = \sum_{i=1}^{r} \Delta_i^{\mathrm{B}}, \quad \text{with } \Delta_i^{\mathrm{B}} \in \mathbb{R}^d, \ r = \operatorname{rank}(\mathcal{U}^{\mathrm{B}}),
\end{equation}
where each \( \Delta_i^{\mathrm{B}} \) lies in a principal direction of the high-rank gradient subspace $\mathcal{U}^{\mathrm{B}}$. Let \( \phi_i \) denote the angle between \( \Delta_i^{\mathrm{B}} \) and the shared vulnerable subspace \( \mathcal{S}(x) \). Then the total projected energy satisfies:
\begin{equation}
    \|\Pi_{\mathcal{S}} \Delta^{\mathrm{B}}\|_2^2 
    \ge 
    \sum_{i=1}^{r} \|\Delta_i^{\mathrm{B}}\|_2^2 \cos^2\phi_i.
    \label{eq_back_projection}
\end{equation}
This bound follows from projection linearity. \revise{The inequality is informative provided the backward gradient components exhibit non-negligible (nonnegative) alignment with the shared subspace; otherwise, the projected energy may vanish despite the backward subspace being high-rank. This alignment condition is empirically validated in \Cref{tab_ablation_backbone}, where adversarially trained backbones exhibit consistently positive gradient similarity across architectures, whereas nonrobust models do not. Consequently, \Cref{lemma_cascade_transfer} does not apply when backward gradients are largely orthogonal to the shared vulnerable subspace, which explains the empirical failure of nonrobust backbones (see \Cref{tab_ablation_backbone}).}

In contrast, the forward update is low-rank, concentrating all energy in a single dominant direction:
\begin{equation}
    \Delta^{\mathrm{F}} = \alpha \cdot u, \quad u \in \mathbb{S}^{d-1},
\end{equation}
where \( u \) is the forward update direction. Therefore, 
\begin{equation}
    \|\Pi_{\mathcal{S}} \Delta^{\mathrm{F}}\|_2^2 
    = \|\Delta^{\mathrm{F}}\|_2^2 \cos^2\theta,
    \label{eq_fore_projection}
\end{equation}
where \( \theta \) is the angle between \( u \) and \( \mathcal{S}(x) \).

\textbf{c. Comparison.}
Putting the two cases together, we see that backward learning distributes its update across $r$ directions, each contributing a non-trivial projection onto the shared subspace $\mathcal{S}(x)$. Even if individual components are only weakly aligned, their aggregate projection grows with $r$, as captured by \Cref{eq_back_projection}. 

By contrast, forward learning concentrates all energy in a single direction, yielding \Cref{eq_fore_projection}. When $f_{\mathrm{back}}$ and $f_{\mathrm{tgt}}$ differ, $\cos^2\theta$ can be small, while the backward average remains significant due to the accumulation of multiple contributing directions. 

We empirically confirm that inequality in \Cref{lemma_cascade_transfer} holds, supported by two key observations in \Cref{tab_2epochs}:
\begin{itemize}
    \item Backward-learning–based defenses consistently outperform forward-learning–based defenses on unseen (transfer) models;
    \item Forward learning overfits its single low-rank direction, since reducing the number of steps mitigates this overfitting and improves its transferability.
\end{itemize}
\end{proof}

\begin{tcolorbox}[takeaway,title=Takeaway]
Forward learning, by enforcing ``being right,'' concentrates its update in a single low-rank direction that may overfit the backbone and thus fail to overlap with the vulnerable subspace of unseen models. In contrast, backward learning, by enforcing ``not being wrong,'' distributes energy across many directions, leading to stronger overlap with the shared subspace and hence higher transferability.
\end{tcolorbox}


\begin{lemma}
\label{lemma_reversion_appx}
Let \( \mathrm{Def}(x) = \mathrm{Def}(x, f_{\mathrm{back}}) \) denote a preemptive defense with random initialization applied using backbone model \( f_{\mathrm{back}} \). Define the first defended image as \( x^{\mathrm{def}} = \mathrm{Def}(x) \), and reapplying this defense using a surrogate backbone \( f_{\mathrm{sur}} \) as \( (x^{\mathrm{def}})^{\mathrm{def}}_\mathrm{sur} = \mathrm{Def}_\mathrm{sur}(x^{\mathrm{def}}) = \mathrm{Def}(x^{\mathrm{def}}, f_{\mathrm{sur}}) \).

In the white-box setting (\( \mathrm{Def} = \mathrm{Def}_\mathrm{sur}\)), such reapplying approximately reconstructs the first perturbations:
\begin{equation}
    (x^{\mathrm{def}})^{\mathrm{def}} - x^{\mathrm{def}} \approx x^{\mathrm{def}} - x.
\end{equation}
In gray- or black-box settings (\( \mathrm{Def} \ne \mathrm{Def}_\mathrm{sur} \)), this approximation deteriorates due to gradient mismatch.
\end{lemma}

Thus, the attacker can reverse the preemptive defense by subtracting the second perturbation:
\begin{equation}
    \tilde{x} = x^{\mathrm{def}} - \left((x^{\mathrm{def}})^{\mathrm{def}}_\mathrm{sur} - x^{\mathrm{def}}\right) = 2x^{\mathrm{def}} - (x^{\mathrm{def}})^{\mathrm{def}}_\mathrm{sur}.
\end{equation}
In the white-box case, this equation gives $\tilde{x} \approx x$, indicating that the clean input is nearly recovered. As shown in \Cref{fig_reversion}, only white-box reversion mostly cancels the defense, restoring perceptual similarity and vulnerability, whereas gray/black-box or purification variants fail due to gradient misalignment, showing the need for full backbone access.

\begin{proof}
Let $x^{\mathrm{def}} = \mathrm{Def}(x) = x + \delta^{\mathrm{def}}(x)$ denote the preemptive defense applied with a random start.

\textbf{a. White-box setting (same backbone, same weights):}
In this ideal setting, the first and reapplied defenses use the exact same model and weights (\( \mathrm{Def} = \mathrm{Def}_\mathrm{sur}\)), and both apply a randomized defense (\ie with random start). Thus, the perturbation \(\delta^{\mathrm{def}}(x)\) is stochastic but drawn from the same distribution \(\mathcal{D}_x\) on each call.

Let \(\delta_1 \sim \mathcal{D}_x\) and \(\delta_2 \sim \mathcal{D}_{x + \delta_1}\) be two independent draws corresponding to the first and second defenses. Then,
\begin{equation}
  \begin{aligned}
    &x^{\mathrm{def}} = x + \delta_1, \quad
    (x^{\mathrm{def}})^{\mathrm{def}} = x + \delta_1 + \delta_2,\\
    &\tilde{x} = 2x^{\mathrm{def}} - (x^{\mathrm{def}})^{\mathrm{def}} = x + \delta_1 - \delta_2.
  \end{aligned}
\end{equation}

Although \(\delta_1\) and \(\delta_2\) are independently sampled, they are generated by the same defense mechanism and model weights, applied to nearby inputs \(x\) and \(x + \delta_1\). Since the defense is typically Lipschitz-continuous in input (following the standard assumption in Theorem~A.1 of~\cite{madry2018towards}) and \(\|\delta_1\|\) is small by design, the distributions \(\mathcal{D}_x\) and \(\mathcal{D}_{x + \delta_1}\) are close in total variation. Hence, the perturbations remain close:
\begin{equation}
    \|\delta_1 - \delta_2\| = \mathcal{O}(L_\delta \|\delta_1\|),
\end{equation}
where \(L_\delta\) is an effective Lipschitz constant of the defense. Thus, the reversion error is:
\begin{equation}
    \|\tilde{x} - x\| = \|\delta_1 - \delta_2\| = \mathcal{O}(L_\delta \|\delta^{\mathrm{def}}(x)\|),
    \label{eq_white_reversion}
\end{equation}
which is small for typical perturbation budgets. Empirically, as shown in Figure 6, we observe that while cancellation is not perfect due to randomness, it is strong: most of the perturbation is neutralized, and \(\tilde{x}\) lies close to the original input. This makes white-box reversion the most effective among all settings.

\textbf{b. Gray-box setting (same architecture, different weights):}
In the gray-box setting, the reapplied defense knows the model architecture but not the exact weights, representing the strongest non-white-box scenario. They construct a surrogate defense:
\begin{equation}
    \mathrm{Def}_{\mathrm{sur}}(x^{\mathrm{def}}) = x^{\mathrm{def}} + \delta^{\mathrm{def}}_{\mathrm{sur}}(x^{\mathrm{def}}),
\end{equation}
where the perturbation is again stochastic but drawn from a different distribution due to model mismatch.

We decompose the expected surrogate perturbation relative to the true one:
\begin{equation}
    \delta^{\mathrm{def}}_{\mathrm{sur}}(x^{\mathrm{def}}) = \alpha^\star \delta^{\mathrm{def}}(x^{\mathrm{def}}) + \beta(x^{\mathrm{def}}),
    \label{eq_reversion_decompose}
\end{equation}
where \(\alpha^\star = \cos\bigl(\delta^{\mathrm{def}}_{\mathrm{sur}}(x^{\mathrm{def}}), \delta^{\mathrm{def}}(x^{\mathrm{def}})\bigr)\) measures average alignment, and \(\beta(x^{\mathrm{def}})\) is the orthogonal residual. Based on the conclusion of the white-box setting that \(\delta^{\mathrm{def}}(x^{\mathrm{def}}) \approx \delta^{\mathrm{def}}(x)\), \Cref{eq_reversion_decompose} can be reformulated as:
\begin{equation}
    \delta^{\mathrm{def}}_{\mathrm{sur}}(x^{\mathrm{def}}) \approx \alpha^\star \delta^{\mathrm{def}}(x) + \beta(x^{\mathrm{def}}).
    \label{eq_reversion_decompose_final}
\end{equation}
Then,
\begin{equation}
  \begin{aligned}
    \tilde{x} &= x^{\mathrm{def}} - \delta^{\mathrm{def}}_{\mathrm{sur}}(x) \\
    &\approx x + \delta^{\mathrm{def}}(x) - (\alpha^\star \delta^{\mathrm{def}}(x) + \beta(x^{\mathrm{def}})) \\
    &\approx x + (1 - \alpha^\star)\delta^{\mathrm{def}}(x) - \beta(x^{\mathrm{def}}),
  \end{aligned}
\end{equation}
with residual norm
\begin{equation}
    \|\tilde{x} - x\| \approx \sqrt{(1 - \alpha^\star)^2 \|\delta^{\mathrm{def}}(x)\|^2 + \|\beta(x^{\mathrm{def}})\|^2}.
\end{equation}
Empirically, robust networks exhibit moderate alignment (\(0 < \alpha^\star < 1\)), so $(1 - \alpha^\star) \|\delta^{\mathrm{def}}(x)\|$ indicates partial cancellation occurs. The improvement over doing nothing is clear, but the recovery is weaker than in the white-box setting. Compared to the white-box case (\Cref{eq_white_reversion}), this residual is strictly larger:
\begin{equation}
    \sqrt{(1 - \alpha^\star)^2 \|\delta^{\mathrm{def}}(x)\|^2 + \|\beta(x^{\mathrm{def}})\|^2} \gg \mathcal{O}(L_\delta \|\delta^{\mathrm{def}}\|) ,
\end{equation}
provided the defense is Lipschitz and the surrogate is imperfectly aligned (\ie \(\alpha^\star < 1\)). 

In contrast to the white-box setting, where the residual shrinks proportionally with the perturbation norm, the gray-box residual includes an orthogonal component \(\|\beta(x^{\mathrm{def}})\|\) that cannot be eliminated through alignment alone. This makes gray-box reversion fundamentally less effective and more distorted.

\textbf{c. Black-box setting (different architecture, zero knowledge):}
Here, the second defense builds a standard surrogate model (\( \mathrm{Def} \ne \mathrm{Def}_\mathrm{sur} \)) with different architecture and no access to the target model’s weights. Prior work \cite{ilyas2019adversarial, tramer2018ensemble} shows that such models learn features that are nearly orthogonal to those used by robust models.

Consequently, the cosine similarity between the true and surrogate perturbations satisfies:
\begin{equation}
    \cos\bigl(\delta^{\mathrm{def}}_{\mathrm{sur}}(x^{\mathrm{def}}), \delta^{\mathrm{def}}(x^{\mathrm{def}})\bigr) \approx 0.
\end{equation}
So the decomposition using \Cref{eq_reversion_decompose_final} becomes:
\begin{equation}
    \delta^{\mathrm{def}}_{\mathrm{sur}}(x^{\mathrm{def}}) \approx \beta(x^{\mathrm{def}}), \quad \alpha^\star \approx 0,
\end{equation}
yielding
\begin{equation}
  \begin{aligned}
    \|\tilde{x} - x\| &\approx \sqrt{ \|\delta^{\mathrm{def}}(x)\|^2 + \|\beta(x^{\mathrm{def}})\|^2} \\
    &> \|\delta^{\mathrm{def}}(x)\|.
  \end{aligned}
\end{equation}
That is, there is \emph{almost no cancellation}. Worse, since the surrogate gradients amplify irrelevant directions, the distortion may even increase, leading to degraded image quality.

\textbf{d. Non-gradient-based ``purifiers":}
Suppose the attacker applies a generic purification operator \(P\), such as JPEG compression or a denoising model, to cancel the defense as an adaptive reversion attack. The result is \(\hat{x} = P(x^{\mathrm{def}})\). These operators are non-invertible, typically satisfying only a contractive bound:
\begin{equation}
    \|P(u) - P(v)\| \le \kappa \|u - v\| \quad (\kappa < 1),
\end{equation}
rather than an inverse identity \(P(x + \delta^{\mathrm{def}}(x)) = x\).

Thus,
\begin{equation}
  \begin{aligned}
    \|\hat{x} - x\| &= \|P(x + \delta^{\mathrm{def}}(x)) - x\| \\
     &\ge (1 - \kappa) \|\delta^{\mathrm{def}}(x)\| - \underbrace{\|P(x) - x\|}_{\mathrm{content loss}}.
  \end{aligned}
\end{equation}
Since \(\delta^{\mathrm{def}}(x)\) often contains low-frequency or semantic components, the purifier does not effectively suppress it. Moreover, the content loss can dominate, further distorting the image. In practice, a single pass of purification fails to reverse the defense and often reduces image quality (see \Cref{fig_reversion}).
\end{proof}

\begin{tcolorbox}[takeaway,title=Takeaway]
\begin{itemize}
    \item \textit{White-box:} stochastic but well-aligned; \(\tilde{x} \approx x\) with small variance — strongest cancellation.
    \item \textit{Gray-box:} moderate alignment leads to partial cancellation; recovery quality depends on \(\alpha^\star\).
    \item \textit{Black-box:} perturbations are nearly orthogonal; cancellation fails, and distortion may increase.
    \item \textit{Non-gradient purifier:} weak and non-specific; typically introduces more distortion than it removes.
\end{itemize}
\end{tcolorbox}

\end{document}